\documentclass[twocolumn]{aastex62}
\usepackage{gensymb}
\usepackage{natbib}
\usepackage{verbatim}
\usepackage{amsmath}

\usepackage{tikz}
\usepackage{pgfplotstable}
\usepackage{pgfplots}
\pgfplotsset{compat=1.13}
\usetikzlibrary{pgfplots.groupplots}

\usetikzlibrary{arrows,3d} 
\usetikzlibrary{patterns}
\usepackage{capt-of}

\usepackage{booktabs} 

\usepackage{algpseudocode}
\usepackage{listings}
\newcommand{\eq}[1]{Equation~(\ref{#1})}

\newcommand{\dee}[1]{\cdot 10^{#1}}
\newcommand{\Rek}{\mathrm{Re}}
\newcommand{\Rem}{\mathrm{Re}_\mathrm{M}}
\newcommand{\Pm}{\mathrm{Pr}_\mathrm{M}}

\newcommand{\Remcr}{\mathrm{Re}_\mathrm{M, crit}}

\newcommand{\meanBB}{\overline{\mathbf{B}}}

\newcommand{\kf}{k_\mathrm{f}} 
\newcommand{\etat}{\eta_\mathrm{t}}

\shorttitle{Large- and small-scale dynamos}
\shortauthors{V\"ais\"al\"a et al.}

\begin{document}

\title{Interaction of large-- and small--scale dynamos in isotropic turbulent flows from GPU--accelerated simulations}

\author[0000-0002-8782-4664]{Miikka S. V\"ais\"al\"a}
\affiliation{Academia Sinica, Institute of Astronomy and Astrophysics, Taipei, Taiwan}

\author{Johannes Pekkil\"a}
\affiliation{Department of Computer Science, Aalto University, Espoo, Finland}

\author[0000-0002-9614-2200]{Maarit J. K\"apyl\"a}
\affiliation{Max Planck Institute for Solar System Research, Justus-von-Liebig-Weg 3, 37077 Göttingen, Germany}
\affiliation{Department of Computer Science, Aalto University, Espoo, Finland}
\affiliation{Nordita, KTH Royal Institute of Technology and Stockholm University, Roslagstullsbacken 23, SE-10691 Stockholm, Sweden}

\author{Matthias Rheinhardt}
\affiliation{Department of Computer Science, Aalto University, Espoo, Finland}

\author[0000-0001-8385-9838]{Hsien Shang}
\affiliation{Academia Sinica, Institute of Astronomy and Astrophysics, Taipei, Taiwan}

\author[0000-0001-5557-5387]{Ruben Krasnopolsky}
\affiliation{Academia Sinica, Institute of Astronomy and Astrophysics, Taipei, Taiwan}

\email{mvaisala@asiaa.sinica.edu.tw, shang@asiaa.sinica.edu.tw}

\begin{abstract}

Magnetohydrodynamical (MHD) dynamos emerge in many different astrophysical situations where turbulence is present, but the interaction between large-scale (LSD) and small-scale dynamos (SSD) is not fully understood. 
We performed a systematic study of turbulent dynamos driven by isotropic forcing in isothermal MHD with magnetic Prandtl number of unity, focusing on the exponential growth stage. Both helical and non-helical forcing was employed to separate the effects of LSD and SSD in a periodic domain. Reynolds numbers ($\Rem$) up to $\approx 250$ were examined and multiple resolutions used for convergence checks. 
We ran our simulations with the \textit{Astaroth} code, designed to accelerate 3D stencil computations on graphics processing units (GPUs) and to employ multiple GPUs with peer-to-peer communication. We observed a speedup of $\approx 35$ in single-node performance compared to the widely used multi-CPU MHD solver \textit{Pencil Code}.
We estimated the growth rates both from the averaged magnetic fields and their power spectra. At low $\Rem$, LSD growth dominates, but at high $\Rem$ SSD appears to dominate 
in both helically and non-helically forced cases. Pure SSD growth rates follow a logarithmic scaling as a function of $\Rem$.
Probability density functions of the magnetic field from the growth stage exhibit SSD behaviour in helically forced cases even at intermediate $\Rem$.
We estimated mean-field turbulence transport coefficients using closures like the second-order correlation approximation (SOCA). They yield growth rates similar to the directly measured ones
and provide evidence of $\alpha$ quenching. 
Our results are consistent with the SSD inhibiting the growth 
of the LSD at moderate $\Rem$, while the dynamo growth is enhanced at higher $\Rem$.

\end{abstract}


\section{Introduction}
\subsection{Astrophysical background}

Nearly all astrophysical objects, ranging from planets, 
accretion disks, stars and galaxies to the intergalactic 
medium, host magnetic fields coherent over the largest
scales of the system. At the same time, the matter in
which the magnetic fields originate, is in a vigorously
turbulent state, and often the driving scale of the
turbulence is at small or intermediate scales with
respect to the system scale 
\citep[see, e.g.,][]{rincon2019}.
Hence, a theoretical 
explanation of how these objects can sustain large-scale 
magnetic fields, driven by small-scale turbulence, is required.
One such theoretical framework is the theory
of $\alpha \Omega$ dynamos, where helical turbulence
together with large-scale non-uniformities in the 
rotation profile
excite magnetic field at the largest 
scales \citep[as originally proposed by][]{Parker55a}.
Here, kinetic helicity is thought to arise from stratification and rotation.
Large-scale dynamos (hereafter LSD) do not necessarily
need rotational non-uniformities, but can also work
solely based on helical turbulence, then denoted as
$\alpha^2$ dynamos 
\citep[see, .e.g., ][]{KR80}.
As the $\alpha$ effect is such a
fundamental building block of LSDs, studying it in
isolation has been a persistent task.
Many details and questions, however,  still remain open, especially at vigorously turbulent regimes, which to reach numerical models still struggle.

Another dynamo instability, namely the fluctuation dynamo (or small-scale dynamo, hereafter SSD) is excited in 
astrophysical flows for magnetic Reynolds
numbers ($\Rem$) exceeding a threshold value, which is thought
to happen in most astrophysical settings 
\citep[as originally proposed by][]{Kazantsev1968}.
SSD generates
random magnetic fields primarily below the scales of the 
forcing, and their growth rate is high, providing a
plausible explanation for magnetic fields in galaxy
clusters, or fluctuating fields seen on
the solar surface.
The latter, however, remains under
some debate, as SSD in a low magnetic Prandtl number ($\Pm$)
environment like the Sun is notoriously hard to excite in numerical
experiments. 

In nearly all astrophysical objects, LSD and SSD instabilities may co-exist. Their interactions,
however, are poorly understood, mainly because it is very
challenging numerically to include them both in one
and the same model, and only quite recently, such
modelling efforts have become feasible. Also, whenever
turbulent enough regimes can be reached, it becomes
very difficult to disentangle the two dynamos, as
also LSD produces fluctuating magnetic fields by
turbulent tangling of the large-scale field.
One of the earliest theoretical scenarios was 
catastrophic quenching of the LSD by the growing
magnetic fluctuations, resulting in the suppression
of the $\alpha$ effect proportional to ${\Rem}^{-1}$,
meaning in practise that no $\alpha$-effect related LSD could be excited in astrophysical objects \citep{CV91}. This is 
now understood to be a special case, detrimentally constrained by magnetic helicity 
conservation, e.g.\ due to closed boundaries, such that
helicity fluxes cannot occur \citep{Bran2005review}. How these fluxes,
which alleviate catastrophic quenching, occur in
cosmic objects, however, is not known in detail. 

It has also been
proposed that SSD can help LSD in shear dynamos
in the absence of the $\alpha$ effect, through 
the so-called
magnetic shear-current effect
\citep[e.g.][]{SB15a}, but its potential
 still remains debated 
\citep[e.g.][]{SMHD}.
\cite{Hotta2016} claimed that, in simulations of turbulent magnetoconvection, the SSD would first suppress LSD at intermediate $\Rem$, but would let it recover at higher $\Rem$. This result was based on measuring the strength of the large--scale field at a few $\Rem$ values, but the diffusion scheme was changed in between the different runs, due to which a straightforward interpretation is difficult.

Many numerical studies have been undertaken  concentrating on LSDs by helical forcing or SSDs by non-helical forcing,
 in the former case also including
unintentionally or intentionally both dynamos.
\citet{Brandenburg2001} demonstrated how helically driven turbulence could give rise to large-scale structures based on non-local interactions at the forcing scale. Their  
simulations employed isotropically 
driven forced turbulence with various 
resolutions and Reynolds numbers, to systematically 
demonstrate the inverse cascade of MHD-turbulence. 
In some of these runs both LSD and SSD were present, but even
though the evolution of mean and fluctuating fields was
monitored separately, no attempt to study the LSD-SSD
interactions was made. 

In the galactic context,
\citet{Gent2013a} studied supernova-driven flows, where the 
magnetic Reynolds number again permitted both dynamos
together. They made an attempt to separate the growth
rates of SSD and LSD by using a Gaussian smoothing 
procedure, but their setup was too complex to derive
any reliable information on the two dynamo processes.
Moreover, their viscosity scheme allowed for 
a spatially varying
$\Pm$, hence the excitation conditions
for the different dynamos were more favourable in regions with hot
gas, in which most likely, most of the SSD action occurred.
Both dynamos together have also been seen in turbulent
convection simulations \cite{KKB08}, and there different
growth rates were detected for mean and fluctuating
magnetic fields, following a $\Rem^{1/2}$ scaling.
\cite{Brandenburg2018} compare the results from 
several studies of helically and non-helically forced
turbulence models, and show that the
growth rate of the SSD is following the same 
scaling. Moreover, in helical system possessing both dynamos,
the same growth rate is observed. Their 
data, however, is sparse, hence 
the $\Rem$-dependence not very certain.

There is a rich literature on SSDs, studied in
isolation in non-helical setups. Most of it concentrates
either on high $\Pm$, relevant for
ISM and intergalactic medium, or low $\Pm$, important for accretion disks and stellar
convection zones. Both regimes are numerically extremely
challenging, as the magnetic diffusivity has to be 
set to values much lower/higher than viscosity, and
resolving such systems numerically is difficult. Here 
we avoid these complications by concentrating
on the regime $\Pm=1$ and make an 
 effort to analyze the interaction of 
LDS and SSD.

The objective of this study is twofold. First, we aim at 
replicating the work of \citet{Brandenburg2001}, 
and then at exploring a
wider range of resolution and $\Rem$. Second, we provide the first 
physical application of the multi-GPU magnetohydrodynamics code \textit{Astaroth} \citep{Astaroth2017, vaisala2017thesis, Pekkila2019} which features novel methods for efficiently calculating high-order finite-difference derivatives, based on large stencils. 

\subsection{Emergence of GPU computation} \label{sec:acbg}

In the last ten years, the emergence of graphics processing units (GPUs) has
enabled several times higher throughput in data-parallel tasks, compared
with central processing units (CPUs) traditionally used in high-performance
computing\footnote{
A Tesla V100-SXM2-32GB GPU provides an arithmetic performance of $7.83$ TFLOPS
(floating-point operations per second) and $863$ GiB/s off-chip memory
bandwidth~\citep{volta-whitepaper, Jia2018}, whereas an Intel Xeon Gold 6230
CPU has the theoretical peak performance of $1.25$ TFLOPS and $131$ GiB/s
bandwidth~\citep{intel-whitepaper}. Therefore a GPU could theoretically
provide roughly $6\times$ improved throughput in data-parallel tasks.
}. GPUs excel in tasks, where the same operation can be
executed on a very high number of data elements in parallel. In contrast to
CPUs, GPUs have been designed to maximize the throughput of memory systems
with the cost of higher memory access latency~\citep{hennessy_computer}.
Therefore, ensuring there is a sufficient amount of parallel work to hide
latencies is critical for obtaining high performance.

GPUs provide an attractive platform for stencil codes, where each grid point in
the problem domain can be updated in parallel. Stencil codes are commonly used
in, for example, finite-difference fluid simulations~\citep{Brandenburg2001} and image processing~\citep{mullapudi_polymage, kelley_halide}. However, following the multi-core revolution and the introduction of highly-parallel accelerators to general-purpose computing, converting existing codes to use all of the capabilities of the hardware has been a significant challenge~\citep{Asanovic2009}. Writing efficient programs for these architectures often requires deep knowledge in their hardware and execution models. 

GPUs are programmed using the stream programming model, where the programmer defines a stream of instructions to be executed in parallel on a multitude of stream processors. Each 
individual stream
can access data from different memory locations or follow different execution paths.

A notable complication in GPU programming is finding efficient caching techniques to reduce pressure onto off-chip memory. As GPUs are capable of high arithmetic throughput, it is paramount to ensure that the stream processors do not become starved of data. This is especially an issue in high-order finite-difference codes, where the ratio of arithmetic operations to bytes transferred is generally low. Because the optimal caching technique depends on several factors, such as the problem size and stencil shape,  the optimal implementation for one workload does usually not carry to another. This presents a major obstacle in studying physical phenomena, or developing new mathematical models by GPUs, as a significant amount of time must be spent on writing, debugging and optimizing the code. 

There have been several proposals to make GPU programming more convenient, e.g. high-level language extensions, such as \textit{OpenACC}~\citep{openacc}. However, high-level programming models are argued to lack the expressiveness to translate more complex tasks, especially those that require advanced caching techniques, into efficient code~\citep{sujeeth_delite, edwards2014}. More specialized approaches have also been suggested. Frameworks focusing on solving PDEs in structured grids include \textit{SBLOCK}~\citep{brandvik_sblock}, \textit{Fargo3D}~\citep{benitez_fargo3d} and \textit{Cactus}~\citep{goodale_cactus}. Alternative approaches focused on achieving near hand-tuned performance and performance-portability have been demonstrated by \textit{Lift}~\citep{hagedorn_high} and \textit{Delite}~\citep{sujeeth_delite}. \textit{Lift} translates high-level algorithmic primitives into lower-level code based on rewrite rules, while \textit{Delite} provides an intermediate language, which can be used as a basis for building domain-specific languages (DSLs). 

Yet another approach is to provide a compiler for generating efficient code
from sources written in a DSL. For example, \textit{Polymage}~\citep{mullapudi_polymage} and \textit{Halide}~\citep{kelley_halide} provide a DSL and a compiler for generating two-dimensional image processing pipelines. In contrast to the above approaches, \textit{Astaroth} has been tailored for high-order stencil computations with special consideration of caching coupled fields 
commonly found in multiphysics simulations,
such as velocity coupled with magnetic field in MHD induction equation (See Equation \ref{eq:magnetic}). 
By caching the results of intermediate stencil operations, traffic to main memory is reduced significantly when these intermediate results are used to update multiple fields. 

In previous work, we have presented the \textit{Astaroth} library~\citep{Pekkila2019}. 
In the case of this study, we have extended it to work on multiple GPUs and utilized it to simulate resistive MHD turbulence and the emergence of dynamos. Inter-GPU communication is carried out using peer-to-peer memory copy functions provided by the CUDA API. However, as these functions do not support inter-node transfers, our implementation is limited to computations within a single node. An implementation for multiple nodes is the subject of ongoing work.

\section{Core methods}\label{sec:astaroth}

In this study, we use the model of magnetohydrodynamics to examine the growth of small- and large-scale dynamos (SSD and LSD respectively).
Hence, the systems examined are essentially non-linear.
To excite a dynamo, turbulence is a highly suitable (but not necessary) ingredient.
Some properties of dynamo-capable systems can be described qualitatively:
In a small-scale dynamo, the magnetic field grows through cascading turbulence and the resulting entanglements, at  scales smaller than the turbulence driving scale. In such a situation merely 
sufficiently high magnetic Reynolds numbers are required \citep{Haugen2004}. In a large-scale dynamo, turbulent helical flows result in growth of magnetic field at scales larger than the  flow scales, in the extreme at the largest scales possible \citep{Brandenburg2001}. 
To understand LSD action, often the perspective of mean-field theory is taken which is discussed  more closely in Section \ref{sec:soca}. 

Because the dynamo processes are non-linear, exploring them requires direct numerical simulations (DNS). 
They can mimic laboratory experiments by studying how the systems react to changing parameters. 
For such DNS, codes are required which can support two central features. The first is resistive MHD, as a dynamo is commonly supposed to be impossible under ideal-MHD conditions 
because magnetic reconnections are thought to be a necessary part of the self-amplification of the magnetic field \citep{rincon2019}.
The second important feature is  a high-order numerical PDE solver to effectively minimize uncontrolled numerical diffusion and to resolve the fine structure of turbulence with high accuracy \citep{axelnum}.

There are many openly accessible codes which meet these requirements, such as the \textit{Pencil Code}~\citep{Pencil2020}. However, the Pencil Code works presently only with traditional CPU-parallelism via MPI communication.
GPU acceleration can significantly reduce the computational costs, and here we demonstrate the use of the GPU code \textit{Astaroth} with similar properties as the \textit{Pencil Code}. \textit{Astaroth} is a software library developed for accelerating stencil computations especially in high-order accurate simulations. In such tasks, \textit{Astaroth} has been shown to to provide higher throughput and energy-efficiency than CPU-based solvers~\citep{Pekkila2019}. The \textit{Astaroth} library consists of an application-programming interface (API), a domain-specific language (DSL), an optimizing compiler that performs source-to-source translation from DSL sources to CUDA kernels, and a toolbox for carrying out common tasks, such as executing reductions on GPUs.

However, this study uses \textit{Astaroth} in a more specific way.
The \textit{Astaroth} API is surrounded by supplementary code, like tools for purposes of input and output, testing, interfacing, data analysis and data processing. 
These additions allow \textit{Astaroth} to be used as a self-sufficient MHD code. As a self-sufficient MHD code, \textit{Astaroth} has following properties, in addition to the general features provided by the library:
\begin{itemize}
    \item Physics (with DSL)
    \begin{itemize}
        \item Continuity equation.
        \item Momentum equations, with full description of viscosity.
        \item Resistive induction equation.
        \item Energy equation in terms of entropy and ideal gas equation of state. (Not used in this study.)
        \item Isotropic random forcing.
    \end{itemize}
    \item Numerical methods
        \begin{itemize}
        \item 6th-order finite difference scheme for calculating derivatives.
        \item 3rd-order 2N-Runge-Kutta time integration.
    \end{itemize}
    \item Auxiliary tools
    \begin{itemize}
        \item Simulation suite for running DNS, which handles the tasks required on the CPU host side. 
        \item Autotest suite to check coherence between GPU and CPU operations.
        \item Limited live rendering features for testing and demonstration purposes. 
        \item Python toolbox for data post-processing, visualization and analysis.
    \end{itemize}

\end{itemize}

What is novel about \textit{Astaroth}, is the versatile GPU implementation. Because \textit{Astaroth} can be directed using the Domain Specific Language (DSL), it is flexible with adding new physics operations additionally to the ones that exist without making demanding case-by-case implementations on the level of CUDA. In the following Section \ref{sec:multigpu}, we will discuss GPU implementation aspect of this work before returning to the physical problem (Sections \ref{sec:physics} and \ref{sec:results}).   

\textit{Astaroth} outputs time series and binary datacubes, which were reduced,
Fourier transformed, imaged and fitted with \textit{Python} using tools from
the \textit{SciPy}, \textit{NumPy} and \textit{Matplotlib} packages
\citep{2020SciPy-NMeth, harris2020array,Hunter:2007}.
Our analysis resulted in several values from performed post-processing, 
and for this, the \textit{Pandas} tool  \citep{reback2020pandas,mckinney-proc-scipy-2010} proved to be useful. \textit{Pandas} made it possible to
store results from several post-processing routines into an extended table. This table could then be accessed to organize and cross-reference our results with relative ease. 
\textit{Paraview} was utilized for 3D visualization \citep{paraview_ref}.

\section{Multi-GPU implementation}\label{sec:multigpu}

Next, we describe our approach of distributing the workload to multiple GPUs on a single computational node.

\subsection{Terminology} 
\label{sec:terminology}

The finite-difference method belongs to the class of stencil schemes, where
data values assigned to points in a structured grid are updated by sampling the
neighborhood of each grid point according to a specific  
pattern, called a {\it stencil}, see Figure~\ref{fig:stencils}. The radius of a symmetric stencil is denoted as $r$, which is the Chebyshev distance in grid indices
from the center point to the furthest points of the stencil. We use the term $k$th-order stencil for a stencil used to calculate derivatives with $k$th-order accurate finite differences.

The number of grid points in the computational domain is denoted by the triple $\mathbf{N} = (N_x, N_y, N_z)$. In this work, we split the grid along the $z$-axis for $p$ devices 
such that the size of the computational domain, local to each device is $\mathbf{n} = (n_x, n_y, n_z) = (N_x, N_y, N_z / p)$. We use the term \textit{device} to refer to a GPU controlled by a \textit{host} CPU.

Some stencil points, required for updating grid points near its boundaries, fall outside the local computational domain. The entirety of those points is called the halo. The total size of the grid, including the halo, is therefore $\mathbf{m} = (n_x + 2r, n_y + 2r, n_z + 2r)$. For $p >= 2$, some of the halo points map to the computational domain of a neighboring device. The data values at these points must be communicated between the contributing devices after each update step. The structure of the grid is visualized in Figure~\ref{fig:halo}. 

To make a distinction between the halo and the area which is subject to boundary conditions (BCs), we    
call the latter the ghost zone. The ghost zone exists for all 
devices assigned to the boundaries of the global computational domain. Our discussion here focuses         
specifically on communicating non-ghost-zone halos between neighboring devices,  
which we refer to as halo exchange. Halo exchange can be easily adapted to       
support periodic BCs by wrapping ghost zones around the global computational domain instead of excluding them from communication. Other BCs     
may require an additional communication step to update the ghost zone. These     
are left out of scope.

\begin{figure}
\centering
    \begin{tikzpicture}[scale=0.5]
    \foreach \i in {-3,...,3}
    {
        \draw[pattern = north west lines, shift = {(\i, 0)}] (0,0) -- (1,0)--(1,1)--(0,1)--cycle;
        \draw[pattern = north west lines, shift = {(0, \i)}] (0,0) -- (1,0)--(1,1)--(0,1)--cycle;
        \draw[pattern = north west lines, shift = {(\i, \i)}] (0,0) -- (1,0)--(1,1)--(0,1)--cycle;
        \draw[pattern = north west lines, shift = {(\i, -\i)}] (0,0) -- (1,0)--(1,1)--(0,1)--cycle;
    }
    \draw[fill = white, shift = {(0, 0)}] (0,0) -- (1,0)--(1,1)--(0,1)--cycle;
    
    \draw[<->, fill = white, shift = {(-4, 1)}] (0, 0) -- (0, 3) node[midway, left] {$r$};
\end{tikzpicture}
\caption{
Two-dimensional cut of a sixth-order  stencil used for computing first- and second-order derivatives in this work. Here $r = 3$.
}
\label{fig:stencils}
\end{figure}
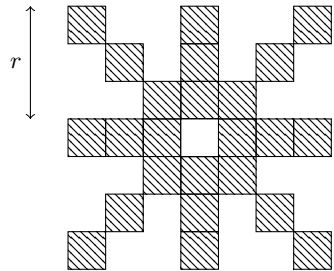

\begin{figure}
\centering
\begin{tikzpicture}[scale=0.35]

    \foreach \y in {-9,...,8}
    {
        \foreach \x in {-9,...,8}
        {
            \draw[pattern = north west lines, shift = {(\x,\y)}] (0,0) -- (1,0)--(1,1)--(0,1)--cycle;
        }
    }

    \foreach \y in {-6,...,5}
    {
        \foreach \x in {-6,...,5}
        {
            \draw[fill=white, shift = {(\x,\y)}] (0,0) -- (1,0)--(1,1)--(0,1)--cycle;
        }
    }

    \foreach \y in {-3,...,2}
    {
        \foreach \x in {-9,...,8}
        {
            \draw[pattern = north east lines, opacity = 0.5, shift = {(\x,\y)}] (0,0) -- (1,0)--(1,1)--(0,1)--cycle;
        }
    }

    \draw[ultra thick] (-9, 0) -- (9, 0);


    \draw[pattern = north east lines](-9, 12) rectangle ++(1,1) node[align = right, anchor = west, midway] {\ Exchanged halo};
    \draw[pattern = north west lines](-9, 11) rectangle ++(1,1) node[align = right, anchor = west, midway] {\ Ghost zone};
    \draw[]                          (-9, 10) rectangle ++(1,1) node[align = right, anchor = west, midway] {\ Computational domain};
    
    \draw[<->, fill = white, shift = {(-10, 6)}] (0, 0) -- (0, 3) node[midway, left] {$r$};
    \draw[<->, fill = white, shift = {(10, 0)}] (0, 0) -- (0, 6) node[midway, right] {$n_z$};
    \draw[<->, fill = white, shift = {(11.5, -6)}] (0, 0) -- (0, 12) node[midway, right] {$N_z$};
    \draw[<->, fill = white, shift = {(-10, -9)}] (0, 0) -- +(0, 12) node[midway, left] {$m_z$};
\end{tikzpicture}
\caption{Visualization of the structured grid used in this work. Each cell represents a grid point.
The thick line represents the separation of the computational domain to two devices.}
\label{fig:halo}
\end{figure}
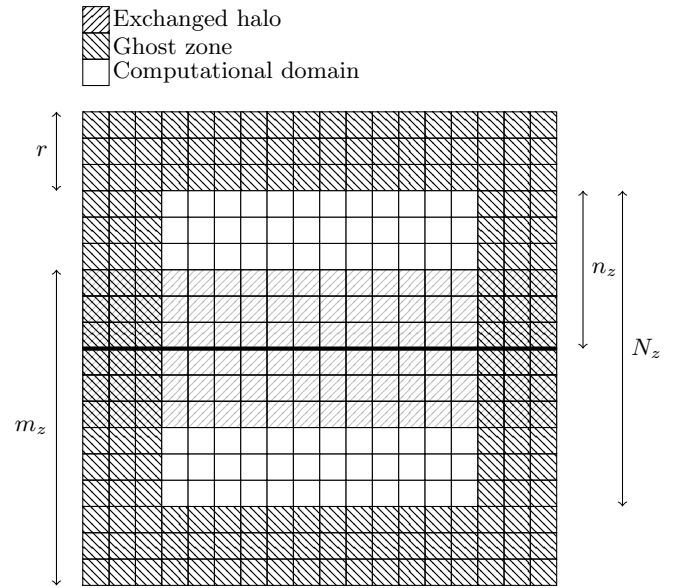

\subsection{Astaroth domain-specific language}
\label{sec:DSL}

The \textit{Astaroth} domain-specific language (DSL) is a stream programming language designed to facilitate the writing of stencil kernels for GPUs. It provides an abstraction level similar to graphics shading languages, such as GLSL~\citep{glsl}. The syntax is an extended subset of C-like languages, providing basic datatypes, operators and tools for control flow, and extending the syntax by adding stream
programming constructs and, for example, function type qualifiers for specifying reusable data. Precision of real numbers is not specified by the DSL, but instead passed as a compilation parameter when building the \textit{Astaroth} executable.

In previous work, we made three assumptions when designing our DSL. Firstly, we assumed that computations are carried out on a structured grid. Secondly, we assumed that each grid point is updated using the same memory access pattern. Finally, we assumed that the result of intermediate operations can be cached and used multiple times when updating a grid point. 
This assumption is the most significant one in terms of performance, as it avoids the 
repetition of expensive intermediate operations which involve reading from slow off-chip memory.
In our case, for example, the current density can be reused to update velocity and entropy. 

\textit{Astaroth} provides a source-to-source compiler for generating efficient CUDA kernels from functions written in the DSL. In addition to optimizations applied during code generation, an automatic optimization is performed at runtime to find the most efficient problem decomposition for the given problem size and hardware.

For this work, we have written the integration kernel used for the MHD simulations solely with the DSL. For further discussion on its syntax  and implementation details, we refer the reader to~\cite{Pekkila2019}.

\subsection{Domain decomposition}

In this work, we decompose the computational domain along a single axis before distributing 
the subdomains to multiple GPUs. The major benefit of this approach is, that it is simple to
implement, while providing sufficiently efficient scaling within a single node. The main 
drawback is, that one-dimensional decomposition is not suitable for large-scale applications, 
due to the fact that the size of the 
exchanged
halo 
decreases at a 
much slower rate as a function of the number of devices compared with multi-dimensional 
decomposition schemes. The benefits of the latter for high-order stencil codes will be discussed in more 
detail in upcoming work.

\subsection{Functions and data dependencies}

In order to hide communication latency, it is critical to carry out computations in parallel with communication. For this, the computational domain must be divided into inner and outer subdomains. The inner computational domain consists of the grid points which can be updated without sampling points in the halo, hence the update can be carried out in parallel with halo exchange. The outer computational domain is formed by the remaining points which can only be updated if the data in the halos is up-to-date. The CUDA API provides concurrency primitives, called streams, which can be used to achieve parallel execution of asynchronous kernels and memory transfers.

We use two buffers for storing the state of the system in order to avoid data races. During integration, we read the data from an input buffer and store the result in a separate output buffer. The buffers are swapped after each substep.

A single simulation step 
comprises the execution of the following functions. Here we use the term \textit{local} to refer to computations or memory operations which do not require 
halo exchange.
The term \textit{global} is used for operations depending on non-local data.
\begin{itemize}
\item \textbf{Local boundary transfer}.  
Update the portion of the halo, which 
depends exclusively on 
data resident in the local memory system
according to the BCs.
\item \textbf{Local update}. Advance the state of the points in the inner computational domain in time.
\item \textbf{Halo exchange}. Exchange a portion of the halos between neighboring devices.
\item \textbf{Global update}. 
Advance the state of the points in the outer computational domain in time.
\item \textbf{Buffer swap}. Swap the input and output buffers in preparation for the next substep.
\item \textbf{Barrier synchronization}. Synchronize the execution state of all devices.
\end{itemize}
The dependencies between these functions and their execution order are visualized in Figure~\ref{fig:concurrency}. Performance enhancements provided by this implementation are listed in Appendix~\ref{sec:performance}. 

\tikzstyle{process} = [rectangle, minimum width = 3cm, minimum height = 1cm, align = center, draw]
\tikzstyle{arrow} = [draw, -latex']
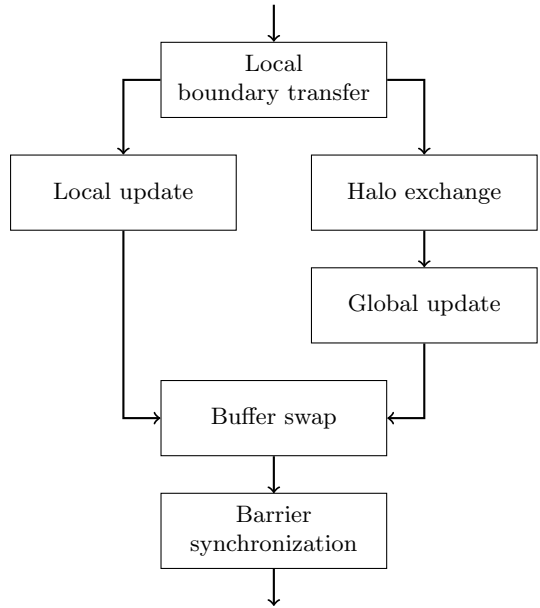
\begin{figure}
    \centering
    \begin{tikzpicture}[node distance = 1.5cm]
    
        \node (local-bound) [process] {Local\\boundary transfer};
        
        \node (local-upd) [process, below of = local-bound, xshift = -2cm] {Local update};
        \node (global-bound) [process, below of = local-bound, xshift = 2cm] {Halo exchange};
        
        \node (global-upd) [process, below of = global-bound] {Global update};
        \node (swap-buffers) [process, below of = global-upd, xshift = -2cm] {Buffer swap};
    
         \node (synchronization) [process, below of = swap-buffers, xshift = 0cm] {Barrier\\ synchronization};
         
         \draw [<-, thick] (local-bound) -- ++(0, 1cm);
        \draw [->, thick] (local-bound) -| (local-upd);
        \draw [->, thick] (local-bound) -| (global-bound);
        
        \draw [->, thick] (global-bound) -- (global-upd);
        \draw [<-, thick] (swap-buffers) -| (global-upd);
        \draw [<-, thick] (swap-buffers) -| (local-upd);
        
        \draw [->, thick] (swap-buffers) -- ++(0, -1cm);
     
         \draw [->, thick] (synchronization) -- ++(0, -1cm);   
    \end{tikzpicture}
    \caption{
        Flowchart of the functions executed during a single integration
        substep. Dependencies are indicated with arrows.}
    \label{fig:concurrency}
\end{figure}

\section{Magnetohydrodynamical model}\label{sec:physics}

We used the continuity, momentum and induction equations of isothermal resistive MHD, corresponding with \citet{Brandenburg2001}:

\begin{equation}\label{eq:continuity}
\frac{D \ln\rho}{D t} = - \nabla \cdot \mathbf{u} 
\end{equation}

\begin{equation}\label{eq:navierstokes}
\begin{split}
\frac{D \mathbf{u}}{D t} = & -c_s^2 \nabla \ln\rho + \frac{\mathbf{j} \times \mathbf{B}}{\rho} \\
& +\nu \bigg[ \nabla^2\mathbf{u} 
+ \frac{1}{3}\nabla(\nabla \cdot \mathbf{u}) 
+ 2 \mathbf{S} \cdot \nabla \ln \rho \bigg] + \mathbf{f}
\end{split}
\end{equation}

\begin{equation}\label{eq:magnetic}
\frac{\partial \mathbf{A}}{\partial t} = \mathbf{u} \times \mathbf{B} + \eta \nabla^2 \mathbf{A},
\end{equation}
where, $\rho$ is density, $\mathbf{u}$ is velocity,
$\mathbf{A}$ is the magnetic vector potential with
$\mathbf{B} = \nabla\times\mathbf{A}$ being the magnetic field, $\mathbf{j} = \nabla\times\mathbf{B}/\mu_0$ is the current density, $\mathbf{S}$ is the traceless rate-of-strain tensor and $\mathbf{f}$ is an external forcing. Of constants, $c_s$ is the isothermal speed of sound, $\mu_0$ the magnetic vacuum permeability, $\nu$ the kinematic viscosity and $\eta$ the ohmic diffusivity. 
Note the use of the {\it diffusive gauge} in \eq{eq:magnetic} for enhanced numerical stability.
In the continuity \eq{eq:continuity} we have used the high-order upwinding method of \citet{Dobler2006} to enhance numerical stability.

We included a forcing function similar to the one in Pencil Code \citep{Pencil2020} 
to generate turbulence. It can be described as, 
\begin{equation}\label{eq:forcing}
\mathbf{f}(\mathbf{x},t) = \mathrm{Re}\bigg\{ N \mathbf{f}_{\mathbf{k}(t)} \exp{\big[i \mathbf{k}(t)\cdot \mathbf{x} + i \phi(t)}\big] \bigg\}.
\end{equation}
Here $\textbf{k}(t) = (k_x, k_y, k_z)$ is a wave vector that changes randomly in each time step, $\textbf{x}$ is a position on the grid and $\phi(t)$ is a random phase in range $[-\pi, \pi]$. The normalization factor is set as
\begin{equation}\label{eq:fnorm}
N = f_0 c_s \sqrt{\frac{k c_s}{\delta t}},
\end{equation}
where $k = |\mathbf{k}|$ and $f_0$ is a scaling factor. For each given time step we randomly generate vectors where $4.5\le|k|\le 5.5$, such that $k_x$, $k_y$ and $ k_z$ are integers. 

We determine eigenfunctions of the curl operator as 
\begin{equation}\label{eq:feigen}
\mathbf{f}_\mathbf{k} = \frac{\mathbf{k} \times (\mathbf{k} \times \mathbf{\hat{e}}) - i \sigma |\mathbf{k}|(\mathbf{k} \times \mathbf{\hat{e}})}{\sqrt{1 + \sigma^2} \mathbf{k}^2 \sqrt{1 - (\mathbf{k}\cdot\mathbf{\hat{e}})^2/\mathbf{k}^2}},
\end{equation} 
where $\mathbf{\hat{e}}$ is a 
random unit vector perpendicular to $\mathbf{k}$.
The forcing function is almost identical to the one presented in \citet{Brandenburg2001}. However, the factor $\sigma \in [0, 1]$ is included to control the degree of helicity, so that with $\sigma = 1$ we get $\sqrt{2}$ instead of $2$ of \citet{Brandenburg2001} in the denominator of \eq{eq:feigen}. Therefore, to match the normalizations, we have set $f_0 = 0.08$ instead of $f_0 = 0.1$ in our models.

For the numerical domain we have adopted the size $L_{x,y,z} = 2\pi$ so that the smallest wave number in the domain is $k_1 = 1$, hence the unit length of $\mathbf{x}$ was set to unity. 
We set the unit of velocity $\mathbf{u}$ to be $c_s = 1$ ,  
and the unit of density $\rho$ to $\rho(\mathbf{x},0)=\rho_0 = 1$, where
$\rho_0$ is the uniform initial density. For the magnetic field $\mathbf{B} $
we choose a unit system in which $\mu_0=1$ and hence set its unit as
$\sqrt{\mu_0\rho_0} c_s = 1$. Therefore the units are equivalent to the ones in
\citet{Brandenburg2001}.

For describing the results, the nondimensional kinetic and magnetic Reynolds numbers,  
\begin{equation}\label{eq:reynolds}
\Rek = \frac{u_\mathrm{rms}}{\nu k_\mathrm{f}} 
\quad \mathrm{and} \quad 
\Rem = \frac{u_\mathrm{rms}}{\eta k_\mathrm{f}} 
\end{equation}
respectively, are useful.
Here $k_\mathrm{f}$ is  the average wave number of the forcing function and $u_\mathrm{rms}$ is the root mean square of the velocity. For the Reynolds numbers presented, we have used measures of $u_\mathrm{rms}$ at the growth stage of dynamo, hereafter $u_\mathrm{rms,0}$. In most of our simulations, $k_\mathrm{f} = 5$ while a few have $\kf=15$, and we keep the magnetic Prandtl number $\Pm=\nu/\eta$  at unity.

Tables \ref{tab:param_LSD} and \ref{tab:param_SSD} list our helical and non-helical turbulence simulation setups. We run our forcing function with both full helicity ($\sigma = 1$) and without helicity ($\sigma = 0$). This allows us to compare effects which are either due to LSD or SSD. We started our simulations with uniform density, zero velocity and a weak Gaussian random magnetic field, which was $\delta$-correlated in space, as a seed field.  We also run our models with multiple resolutions to monitor convergence.

Apart from the forcing helicity, the other physical parameters varied were $\nu=\eta$. At the high end, they correspond to the ones featured in \citet{Brandenburg2001}. At the low end however, the limit for $\nu$ and $\eta$ and therefore $\Rem$ is set by the maximum available resolution, limited by the available total GPU memory in a computational node. A single node with 4 Tesla P100 devices was able to support $512^3$ grid resolution at maximum. A significant part of the analysis was dependent on global averages calculated during runtime, and the reduction method required a computational domain resolution with $2^n$ grid points. 
The version of \textit{Astaroth} used for this work did not yet provide support for multiple nodes, which would have overcome this issue. 

The runs with insufficient resolution tended to crash very early. However, those that kept stable after beginning would keep stable until the end. We avoided using $512^3$ resolution unless necessary to save hard drive space in the computing cluster and for avoiding unnecessary post-processing time. In addition we set $t_\mathrm{max} = 2000$ instead of $4000$ for $512^3$, which might affect our estimates related to the saturation stage. In addition, we run a smaller set of runs to $t_\mathrm{max} = 600$ and a short snapshot interval to perform some more advanced analysis at the exponential growth stage.

During the growth stage, we fit an exponential function to estimate the growth rate $\lambda$ or $\lambda_k$ of our models
\begin{equation}
   B_\mathrm{rms} \propto \exp{(\lambda t)}
   \quad \mathrm{or} \quad 
   E_{B, k} \propto \exp{(2\lambda_k t)},
\end{equation}
depending on whether we estimate the growth rate of a global average (like the rms value) or of a  spectral channel 
of $\mathbf{B}$, $E_{B, k}= B_k^2/2\mu_0$, where $k$ is the wavenumber of the channel. The factor of 2 is required for both $\lambda$ or $\lambda_k$ to agree.
Error estimates are based on Equation (10) of \citet{Morr2014}.

For $\lambda$ we set the time ranges for the fitting by hand, whereas for $\lambda_k$ we find them by fitting in multiple ranges and picking the range with smallest error.
Effectively the fitting errors are negligible, but $\lambda_k$ can have some uncertainties due to automatically picked time ranges.
In Section \ref{sec:soca} we show results for $\lambda$ based on estimated mean-field turbulent transfer coefficients.

\begin{deluxetable}{rccccc}
\tablecaption{Simulation setups and their properties, with 
helical forcing ($\sigma = 1$) \label{tab:param_LSD}}
\tablewidth{0pt}
\tablehead{
$N$ &  $\eta$ &  $\lambda$ &  $\Rem$ &  $\alpha$ & $\etat$ 
}
\startdata
         64 & 5.00e-03 &    1.73e-02 &    3.60e+00 & 2.29e-02 & 4.68e-03 \\
        128 & 5.00e-03 &    1.85e-02 &    3.61e+00 & 2.28e-02 & 4.66e-03 \\
        256 & 5.00e-03 &    1.79e-02 &    3.60e+00 & 2.27e-02 & 4.65e-03 \\
         64 & 3.00e-03 &    2.82e-02 &    7.74e+00 & 2.57e-02 & 5.24e-03 \\
        128 & 3.00e-03 &    2.90e-02 &    7.70e+00 & 2.59e-02 & 5.29e-03 \\
        256 & 3.00e-03 &    2.85e-02 &    7.69e+00 & 2.58e-02 & 5.26e-03 \\
         64 & 2.00e-03 &    3.11e-02 &    1.37e+01 & 2.83e-02 & 5.73e-03 \\
        128 & 2.00e-03 &    3.08e-02 &    1.37e+01 & 2.84e-02 & 5.74e-03 \\
        256 & 2.00e-03 &    3.12e-02 &    1.36e+01 & 2.84e-02 & 5.75e-03 \\
         64 & 1.50e-03 &    2.78e-02 &    1.93e+01 & 3.00e-02 & 6.07e-03 \\
        128 & 1.50e-03 &    2.90e-02 &    1.94e+01 & 3.01e-02 & 6.48e-03 \\
        256 & 1.50e-03 &    2.90e-02 &    1.95e+01 & 3.05e-02 & 6.11e-03 \\
         64 & 1.00e-03 &    2.96e-02 &    3.14e+01 & 3.21e-02 & 6.50e-03 \\
        128 & 1.00e-03 &    2.95e-02 &    3.16e+01 & 3.31e-02 & 6.57e-03 \\
        256 & 1.00e-03 &    3.09e-02 &    3.10e+01 & 3.31e-02 & 6.58e-03 \\
        128 & 7.50e-04 &    3.04e-02 &    4.21e+01 & 3.46e-02 & 6.85e-03 \\
        256 & 7.50e-04 &    3.20e-02 &    4.23e+01 & 3.47e-02 & 6.84e-03 \\
        128 & 5.00e-04 &    3.17e-02 &    6.56e+01 & 3.63e-02 & 7.19e-03 \\
        256 & 5.00e-04 &    3.24e-02 &    6.56e+01 & 3.67e-02 & 7.27e-03 \\
        256 & 4.00e-04 &    3.46e-02 &    8.15e+01 & 3.76e-02 & 7.43e-03 \\
        256 & 2.50e-04 &    3.54e-02 &    1.31e+02 & 3.80e-02 & 7.60e-03 \\
        512 & 2.50e-04 &    3.55e-02 &    1.30e+02 & 3.67e-02 & 7.51e-03 \\
        512 & 2.00e-04 &    3.76e-02 &    1.62e+02 & 3.70e-02 & 7.55e-03 \\
        512 & 1.50e-04 &    4.09e-02 &    2.20e+02 & 3.69e-02 & 7.63e-03 \\
        512 & 1.25e-04 &    4.40e-02 &    2.66e+02 & 3.62e-02 & 7.53e-03 \\
\enddata
\end{deluxetable}

\begin{deluxetable}{rccc}
\tablecaption{Simulation setups and their properties, with non-helical forcing ($\sigma = 0$).  \label{tab:param_SSD}}
\tablewidth{0pt}
\tablehead{
$N$ &  $\eta$ &  $\lambda$ &  $\Rem$ 
}
\startdata
         64 & 5.00e-03 &   -1.35e-02 &    3.53e+00  \\
        128 & 5.00e-03 &   -1.37e-02 &    3.52e+00  \\
        256 & 5.00e-03 &   -1.43e-02 &    3.52e+00  \\
         64 & 3.00e-03 &   -1.65e-02 &    7.16e+00  \\
        128 & 3.00e-03 &   -1.56e-02 &    7.14e+00  \\
        256 & 3.00e-03 &   -1.62e-02 &    7.16e+00  \\
         64 & 2.00e-03 &   -1.22e-02 &    1.21e+01  \\
        128 & 2.00e-03 &   -1.15e-02 &    1.21e+01  \\
        256 & 2.00e-03 &   -1.09e-02 &    1.21e+01  \\
         64 & 1.50e-03 &   -6.70e-03 &    1.73e+01  \\
        128 & 1.50e-03 &   -7.33e-03 &    1.73e+01  \\
        256 & 1.50e-03 &   -5.64e-03 &    1.73e+01  \\
         64 & 1.00e-03 &    1.60e-03 &    2.80e+01  \\
        128 & 1.00e-03 &    1.64e-03 &    2.79e+01  \\
        256 & 1.00e-03 &    2.04e-03 &    2.79e+01  \\
        128 & 7.50e-04 &    7.38e-03 &    3.87e+01  \\
        256 & 7.50e-04 &    7.50e-03 &    3.86e+01  \\
        128 & 5.00e-04 &    1.51e-02 &    6.05e+01  \\
        256 & 5.00e-04 &    1.46e-02 &    6.00e+01  \\
        128 & 4.00e-04 &    1.74e-02 &    7.65e+01  \\
        256 & 4.00e-04 &    1.96e-02 &    7.62e+01  \\
        256 & 2.50e-04 &    2.71e-02 &    1.24e+02  \\
        512 & 2.50e-04 &    2.87e-02 &    1.25e+02  \\
        256 & 2.00e-04 &    3.14e-02 &    1.56e+02  \\
        512 & 2.00e-04 &    3.27e-02 &    1.57e+02  \\
        512 & 1.50e-04 &    3.92e-02 &    2.10e+02  \\
        512 & 1.25e-04 &    4.30e-02 &    2.53e+02  \\
\enddata
\end{deluxetable}

\begin{figure*}[htb!]
\plotone{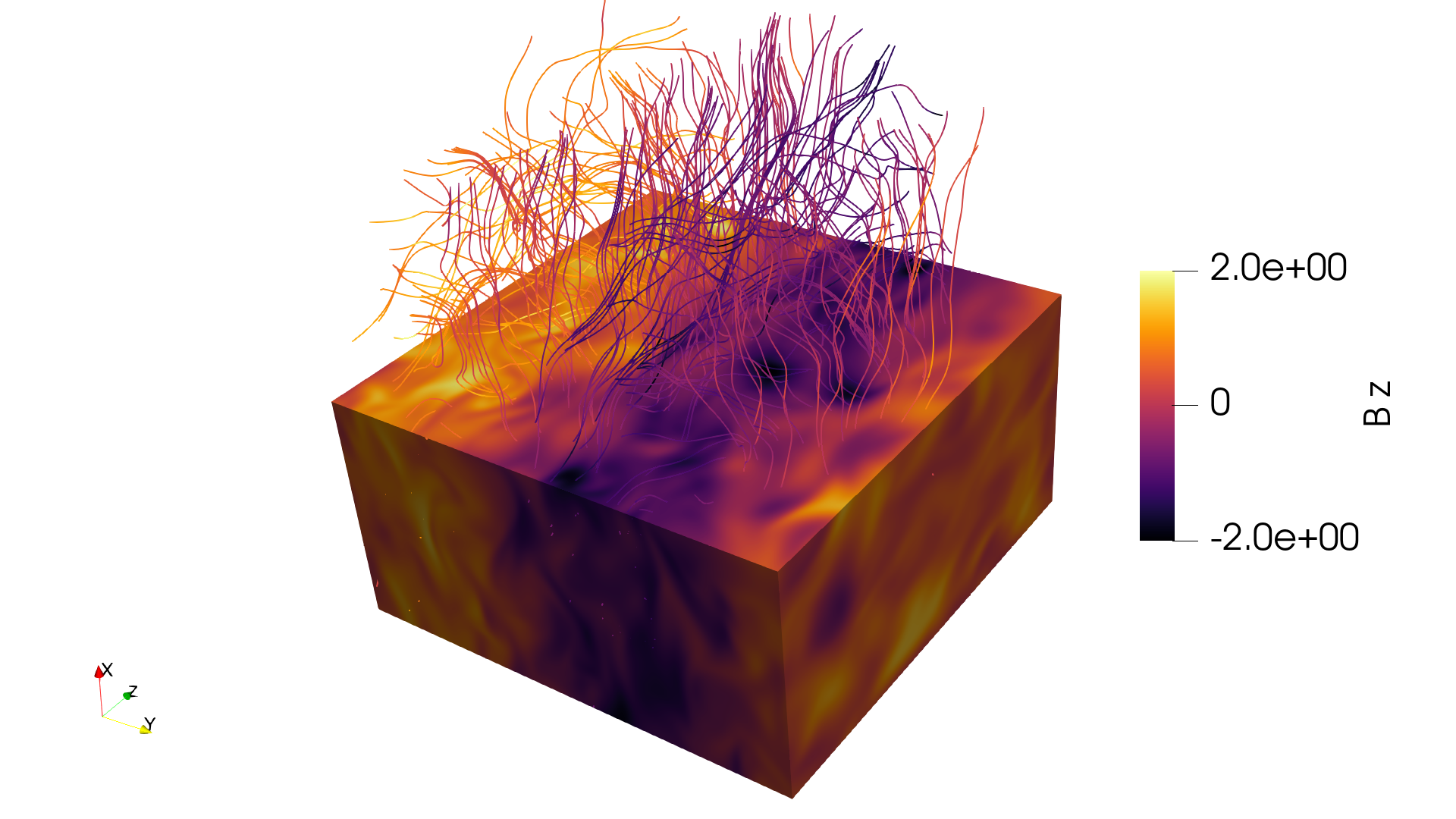}
\plotone{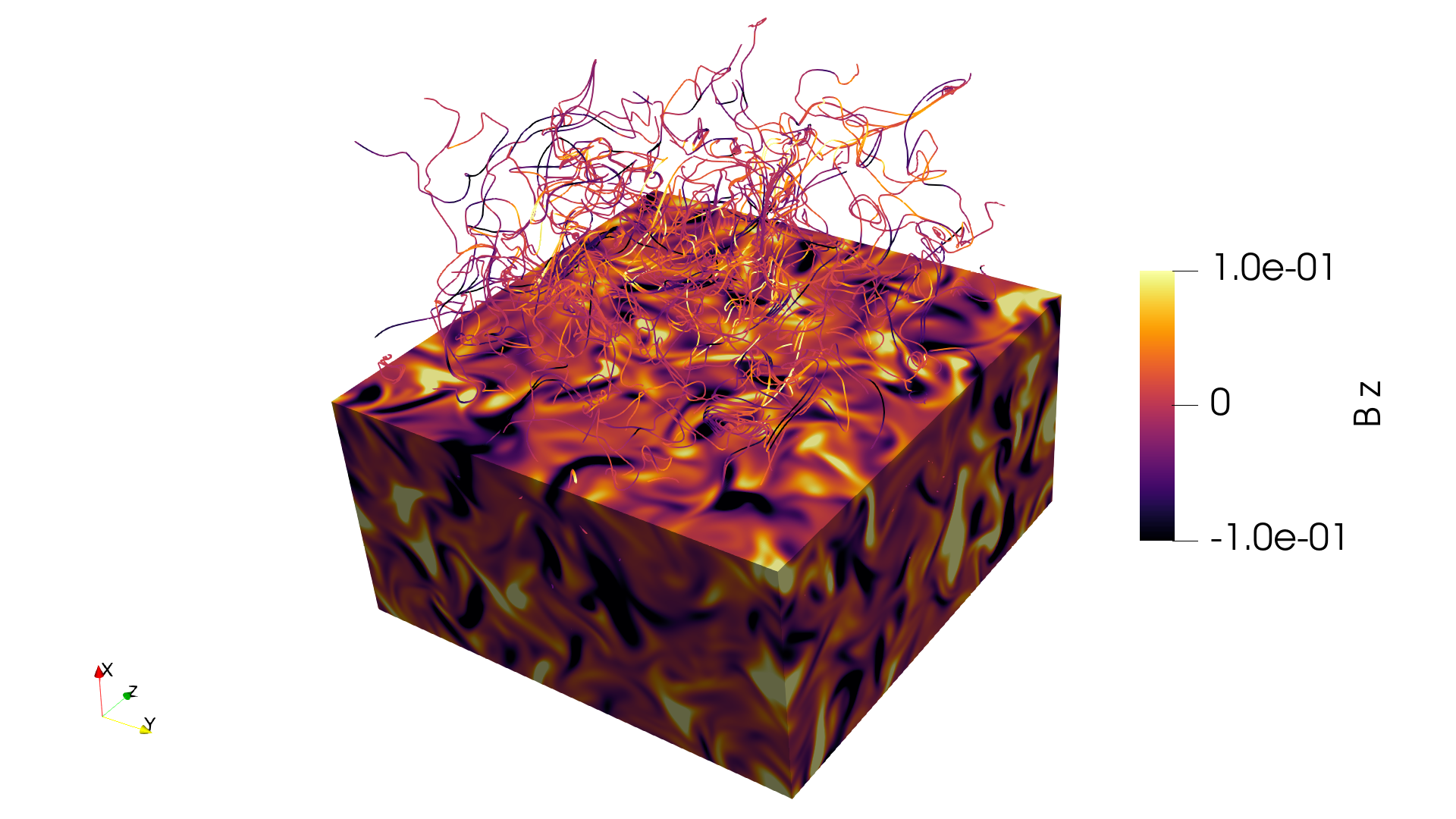}
\caption{Snapshots of dynamo fields in helically (top) and non-helically
(bottom) driven system with $\eta = \nu = 1\dee{-3}$ and $N=256$. Colours
represent $B_z$, normalized with the  equipartition magnetic field
$B_\mathrm{eq} = \sqrt{\mu_0\rho_0} u_\mathrm{rms,0}$. (Animated figures
display growth of the magnetic field from the initial seed field. 
During the early phases of evolution, magnetic fluctuations grow quickly. In
the helical case (top) dominant large-scale structure with $k/k_1=1$ grows more
slowly, but eventually dominates, whereas non-helical case (bottom)
fluctuations grow without large-scale structure.) \ \label{fig:3deta1em3}}
\end{figure*}

\begin{figure*}[htb!]
\plotone{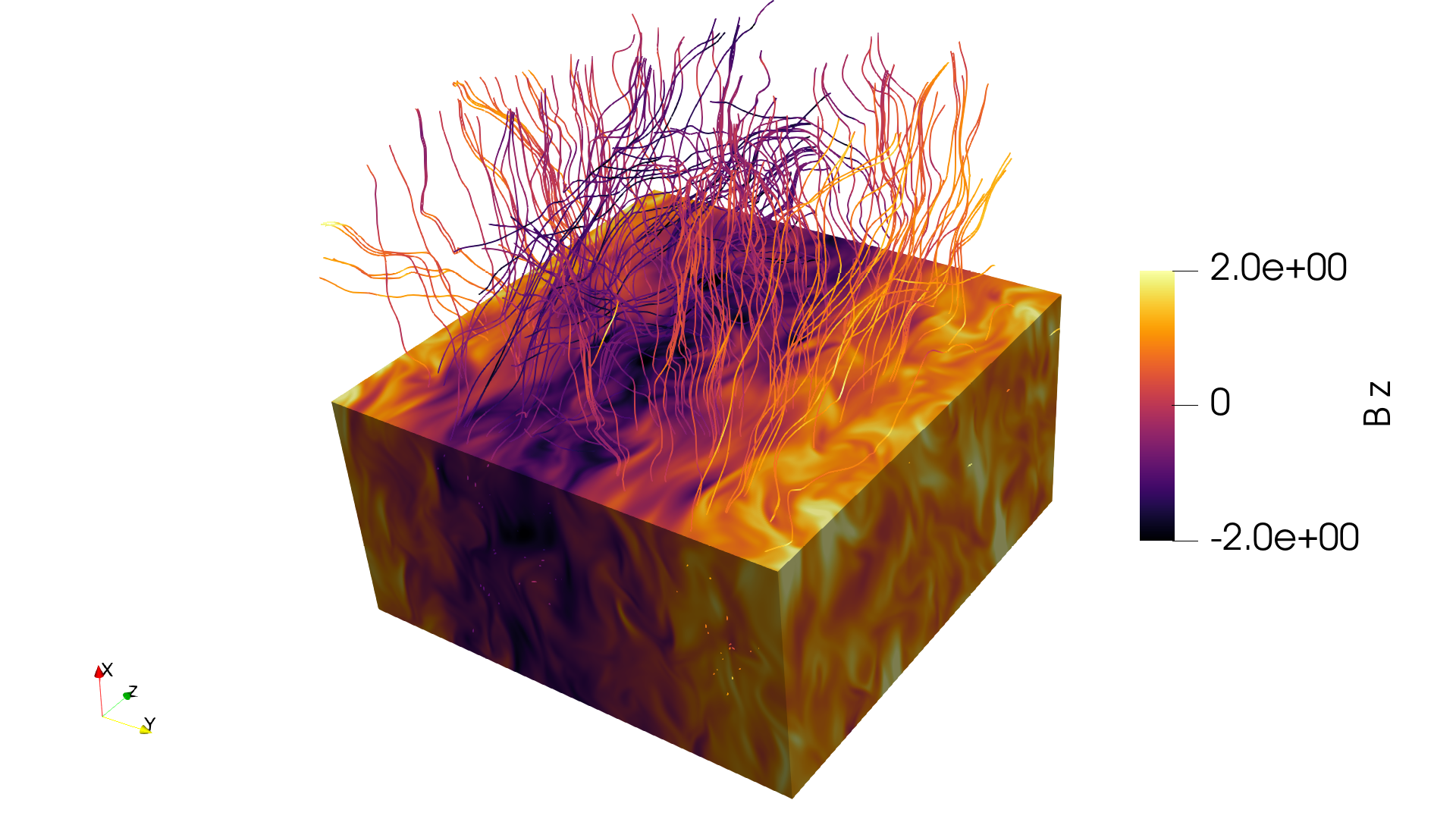}
\plotone{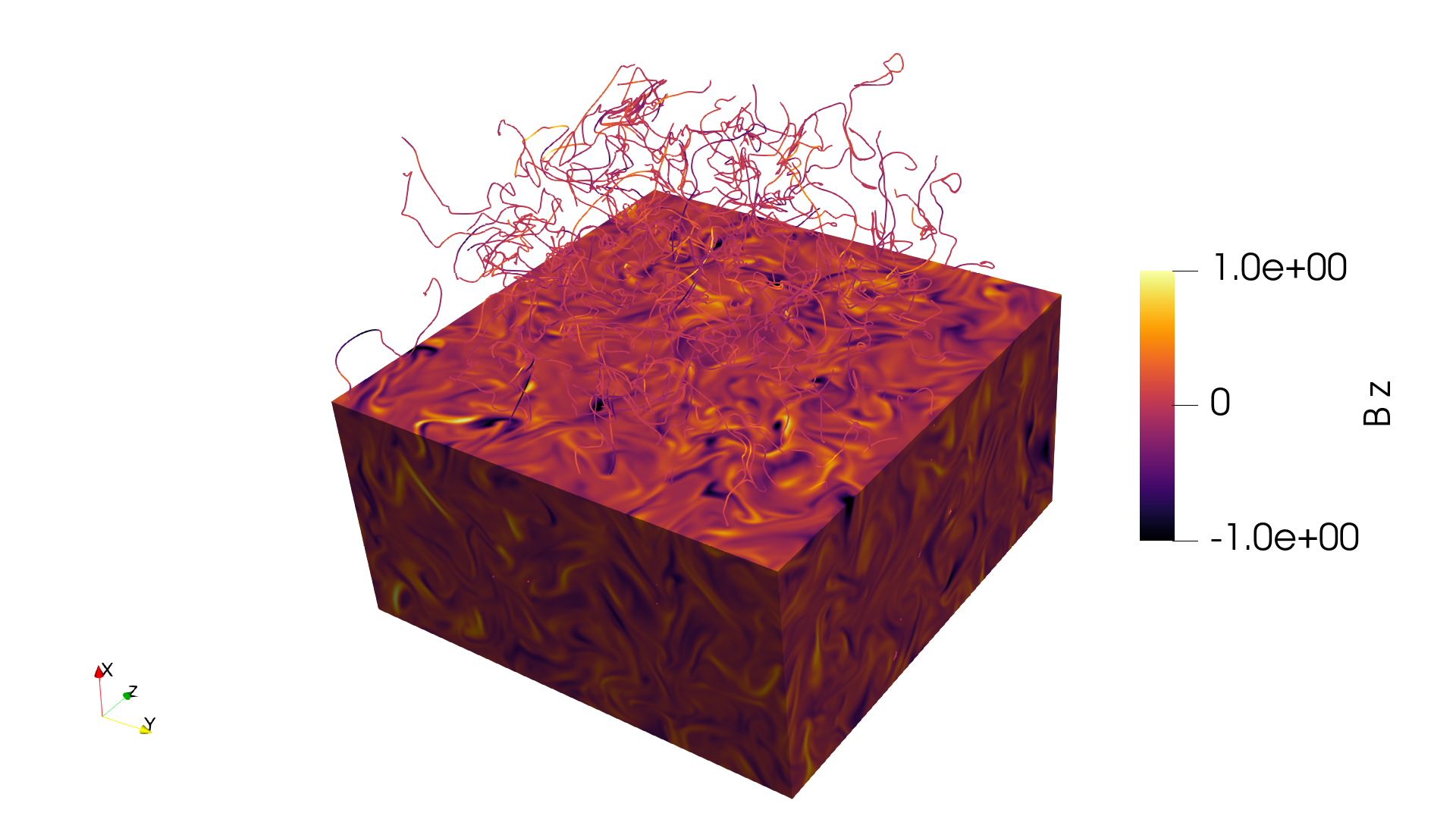}
\caption{As Figure \ref{fig:3deta1em3}, but with $\eta = \nu = 2.5\dee{-4}$. (Animations behave as in Figure \ref{fig:3deta1em3} but turbulent 
fluctuations cascade into smaller scales.) \ \label{fig:3deta25em4}}
\end{figure*}

\section{Results} \label{sec:results} 

\begin{figure*}[ht!]
\plotone{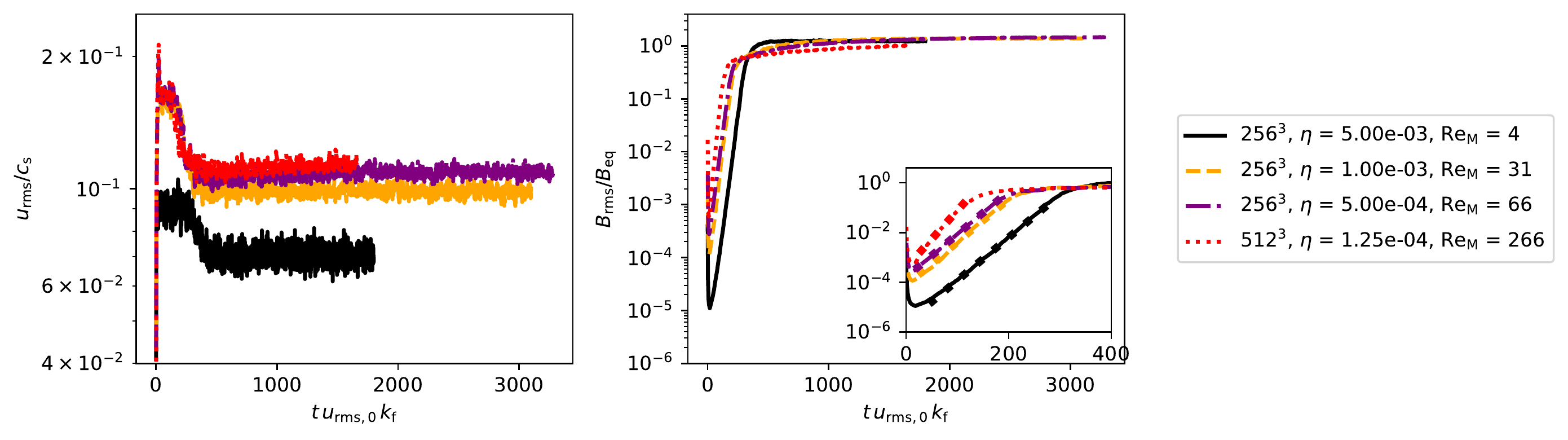}
\caption{Behaviour of different helical setups as a function of time. $u_\mathrm{rms,0}$ is the time average of $u_\mathrm{rms}$ during the expotential growth stage. 
Thick dotted lines within the insets display the exponential fits.
\label{fig:timeseries}}
\end{figure*}

The time development of the runs with helical forcing (see Table \ref{tab:param_LSD}; Figures \ref{fig:3deta1em3}, \ref{fig:3deta25em4} and \ref{fig:timeseries}) exhibits three stages: initial decay, exponential growth, and possible slow growth leading to saturation. 
The initial transient growth of the velocity field, and the contemporary decay of the initial magnetic field, are short for all runs. It is followed by saturation of the rms velocity and exponential growth of the magnetic field. The growing magnetic field starts eventually to quench the velocity, when their energy densities become comparable. The rms velocities then saturate at lower levels, the quenching being the strongest and taking place most slowly the lower $\Rem$ is.
The rms velocities used for calculated $\Rem$ have been measured from the saturated values before quenching starts. Magnetic field values presented in the figures have been normalized with the equipartition magnetic field $B_\mathrm{eq} = \sqrt{\mu_0\rho_0} u_\mathrm{rms,0}$.

For helical forcing, the dynamo will always exhibit a large-scale ($k=1$) magnetic field. 
If there is no simultaneous SSD operating, this field is
well visible during exponential growth and saturation, otherwise it is
fully emerging only during the saturated stage, with weak  signatures  during exponential growth.
Typical field geometries are shown in Figures \ref{fig:3deta1em3} and \ref{fig:3deta25em4} for low and high $\Rem$, respectively. The presence of LSD 
is
expected because in the helically turbulent regime,
the critical dynamo number is close to unity. The emerging large-scale field is of the form of a Beltrami field,
\begin{equation}
    \mathbf{B}\big(\mathbf{x}) = (B_x \sin (k z + \phi), B_y \cos( k z + \phi), 0\big),
\end{equation}
for alignment along $z$, analogously for $x$ and $y$;
$\phi$ is an arbitrary phase.
Alignment and phase are  unpredictable as due to the non--linear nature of the MHD system, even tiny differences in initial conditions or round--off errors may lead to different 
orientation of the Beltrami field being realized in the simulation.

\begin{figure*}[ht!]
\plotone{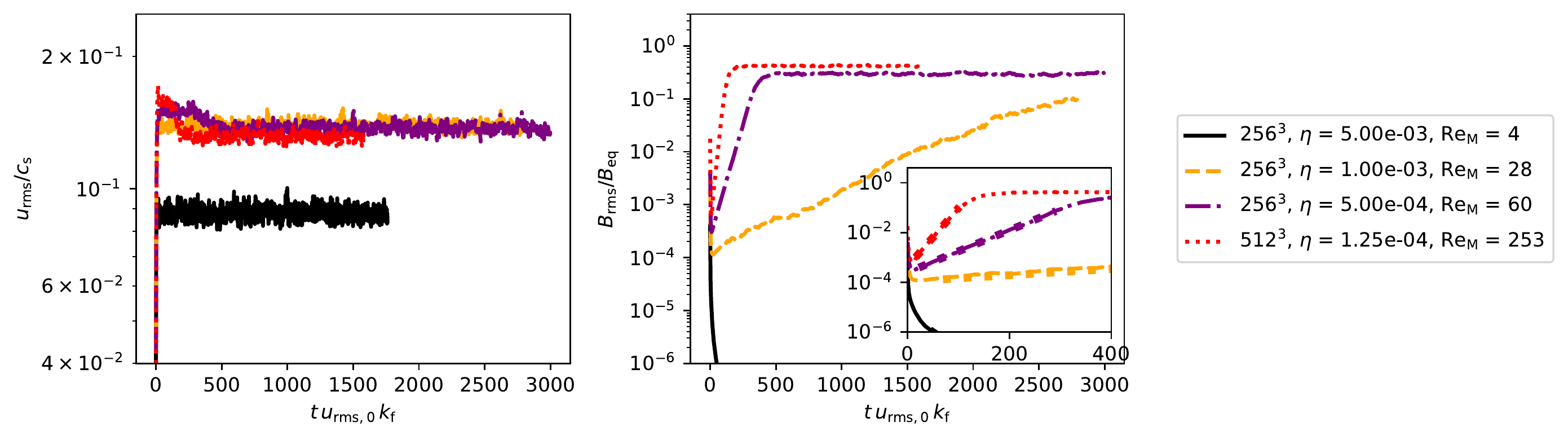}
\caption{As Figure \ref{fig:timeseries}, but for non-helical forcing.  \ \label{fig:timeseries_nohel}}
\end{figure*}

For non-helical forcing, magnetic field growth is not seen for all of our $\Rem$ values, but only above a critical value $\Remcr \sim 25$.
Below it, the magnetic field decays exponentially. In the case of non-helical turbulence, the 
overall structure and geometry of the magnetic field at the saturation stage retains similar form to the growth stage, 
but the magnetic field strength no longer increases (such as Figure \ref{fig:3deta25em4}, bottom).
After the exponential growth, if an LSD is present, there can be still gradual growth of the magnetic field until full saturation is reached. 

At low $\Rem$ where our diffusivity parameters are within the same range as theirs, our results generally agree qualitatively with \citet{Brandenburg2001} with the emergence of $k = 1$ Beltrami field, 
however we do get generally weaker growth rates than they do for unidentifiable reason, with theirs being $\sim 2-3 \times$ larger with the  points having comparable magnetic diffusivity \citep[][Run 2 and Run 3]{Brandenburg2001}. 
-- and at higher $\Rem$ similar principles apply, with the large-scale magnetic field forming, despite increased randomness at smaller scales. 

However, despite the chaotic nature of the system, the resolution does not appear to make a significant difference. As long as $\Rem$ is not too high for the given resolution, practically identical results are produced.
Therefore, for the figures we have chosen representative samples from the highest resolution runs. Agreement across resolutions also indicates that any effects caused by the numerical grid are not significant.

Growth and self-organization of the magnetic field can be seen at different resolutions in the animated Figures \ref{fig:3deta1em3} and \ref{fig:3deta25em4}. With increasing $\Rem$ the small-scale substructures accompanying the coherent mean-field tend to get finer. As we further discuss in Section \ref{sec:spectra}, this could be an indication that an SSD is acting in parallel with the LSD.

\subsection{Growth rates} \label{sec:growth} 

Figures \ref{fig:timeseries} and \ref{fig:timeseries_nohel} show the early exponential growth stage
as well as saturation, cf.
 \citet{Brandenburg2001} (their Fig. 1) for helical turbulence and \citet{Haugen2004} (their Fig. 6) for non-helical turbulence. 
 
\begin{figure}[ht!]
\plotone{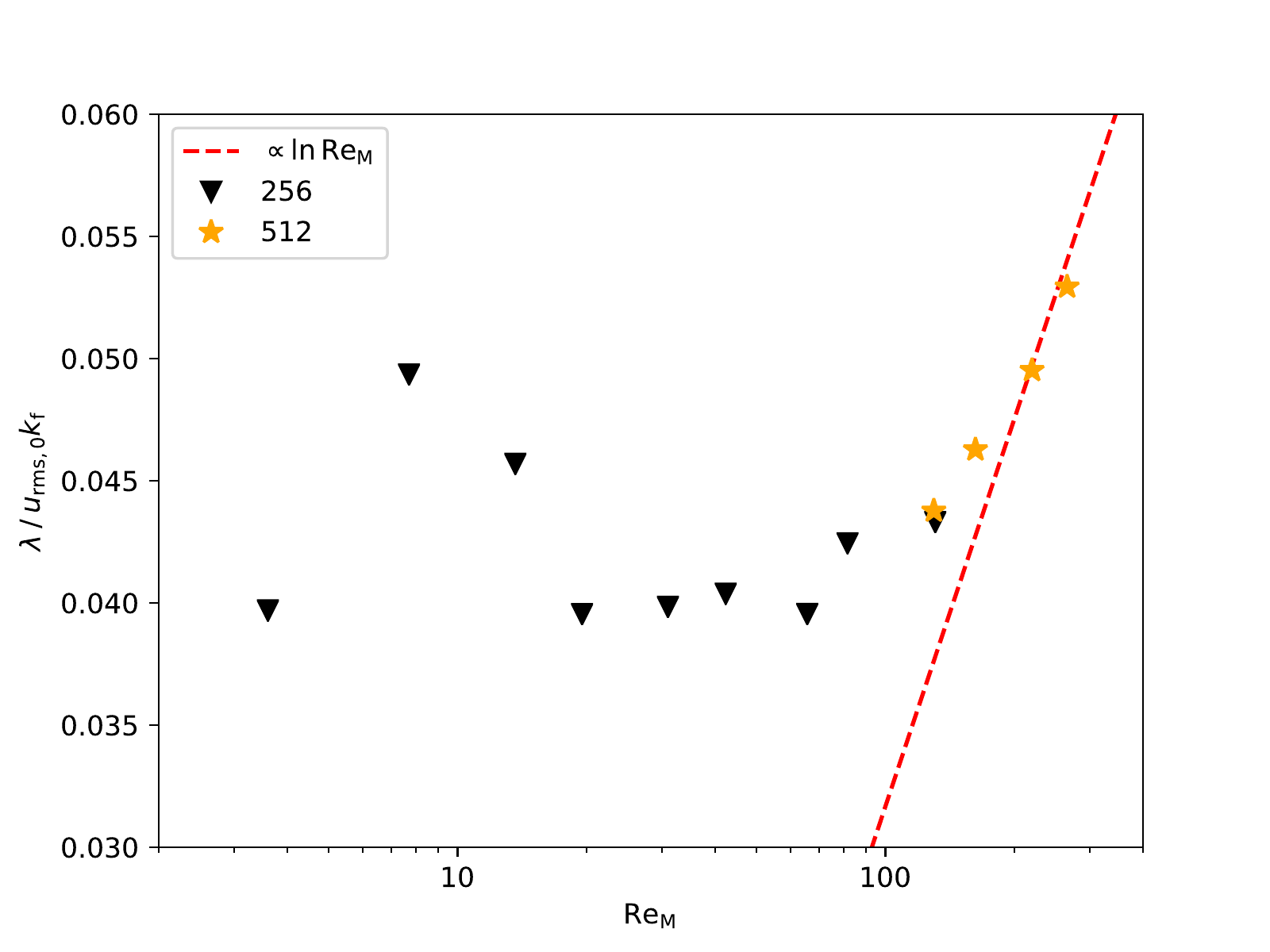}
\caption{Growth rate as a function of magnetic Reynolds number for helical forcing. \label{fig:growthrate}}
\end{figure}

\begin{figure}[ht!]
\plotone{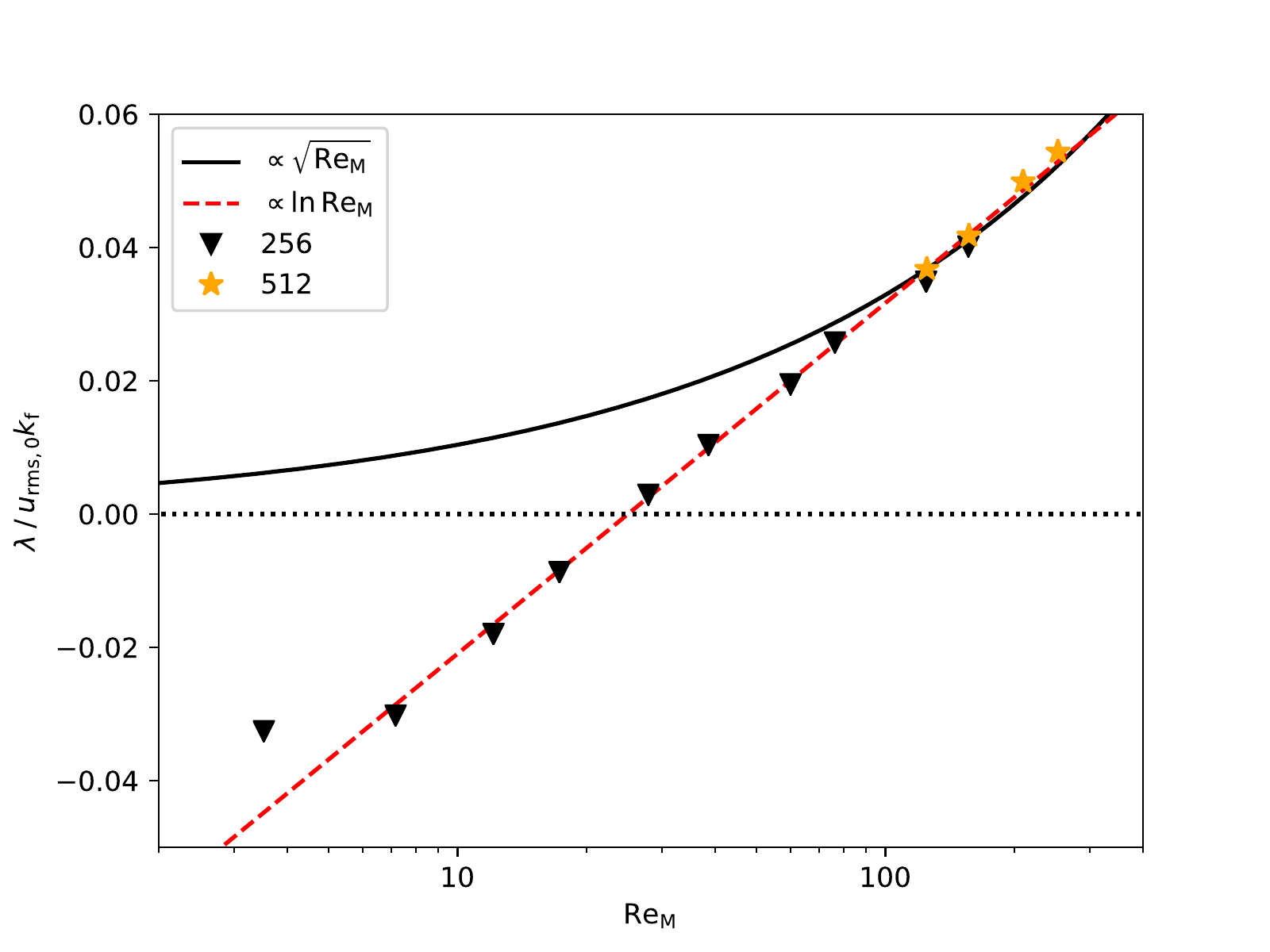}
\caption{Growth rate as a function of magnetic Reynolds number for non-helical forcing. \label{fig:growthrate_nohel} }
\end{figure}

 Our estimated growth rates are shown in Figures \ref{fig:growthrate} and \ref{fig:growthrate_nohel}. 
Those of the helically forced simulations with LSD are positive in all cases. Mildly higher values appear for $\Rem=7.7,13.7$, 
but otherwise the normalized growth rate curve appears flat until $\Rem \approx 100$. 
For the two highest values of $\Rem$, however, the growth rates become similar to those of the non-helical cases. This is indicative of simultaneous SSD action. 
Our results 
agree reasonably well with \citet[Figure 3]{Brandenburg2009}, where they display growth rates at various $\Rem$ for helical forcing, combined with data from \citet{Haugen2004} for the non-helical forcing. Their helical growth rate curve is also flat at low $\Rem\le 70$, but approaches the non-helical growth rates at $\Rem=670$. The helical and non-helical growth rates shown in \citet{Brandenburg2009} align with comparable numerical range to ours. 
However, the comparison can be problematic for two reasons. First, their results are more limited with the respect to the number of data points at with 3 points for helical and 4 for non-helical turbulence. 
Therefore they did not truly resolve the shape of the curve. 
Second, $\Rem$ is varied but $\Rek$ is kept the same leading to variable $\Pm$, which means that our setups are not completely equal type.

\citet{Brandenburg2009} suspected that there is a point where a system would switch from exhibiting merely a LSD to a dynamo combined of LSD and SSD. Our results clearly support this interpretation
as can been seen when comparing Figures \ref{fig:growthrate} and \ref{fig:growthrate_nohel} (see also the power spectra based growth rates in Figure \ref{fig:growthrate_channel} and magnetic field distributions in Figure \ref{fig:BPDF}). It should be noted that if the basic phenomenon observed by \citet{Brandenburg2009} is the same as we observe, this could imply that it is more dependent on  $\Rem$ than $\Rek$. However, our results cannot substantiate this claim, as we only examine $\Pm = 1 $ regime. 

There are two analytical predictions for the SSD growth rate: the more common $\sqrt{\Rem}$ scaling \citep[see e.g.][]{Haugen2004} in contrast to the logarithmic scaling $\ln (\Rem/\Remcr)$, where $\Remcr$ is the critical Reynolds number for the SSD \citep{Kleeorin2012}, and validity is restricted to low magnetic Prandtl numbers and $\Rem \approx \Remcr$. 
In Fig. \ref{fig:growthrate_nohel}, the SSD growth rates appear to be highly consistent with the logarithmic scaling, except at the lowest $\Rem$. 
In contrast, the $\sqrt{\Rem}$ scaling does not really apply, apart from high $\Rem$. 
From Figure \ref{fig:growthrate_nohel}, $\Remcr\sim 25$ has been estimated, while \citet[Fig. 1]{Haugen2004} provide $\sim 35$. 

However, given the uncertainty due to their low number of $\Rem$ data points, 
the estimates might not be significantly different. Figure 2 of \citet{Iskakov2007} shows that for 
incompressible turbulence with $\Pm=1$,
$\Remcr \approx 60$ based on $k_1 = 2\pi$,
or $\approx 42$ if their $\Rem$ is scaled with $\kf/k_1 = \sqrt{2}k_1/k_1 = \sqrt{2}$
instead, as noted by \citet{Brandenburg2018}. 

\subsection{Saturation}
\label{sec:saturation}

\begin{figure}[htb!]
\plotone{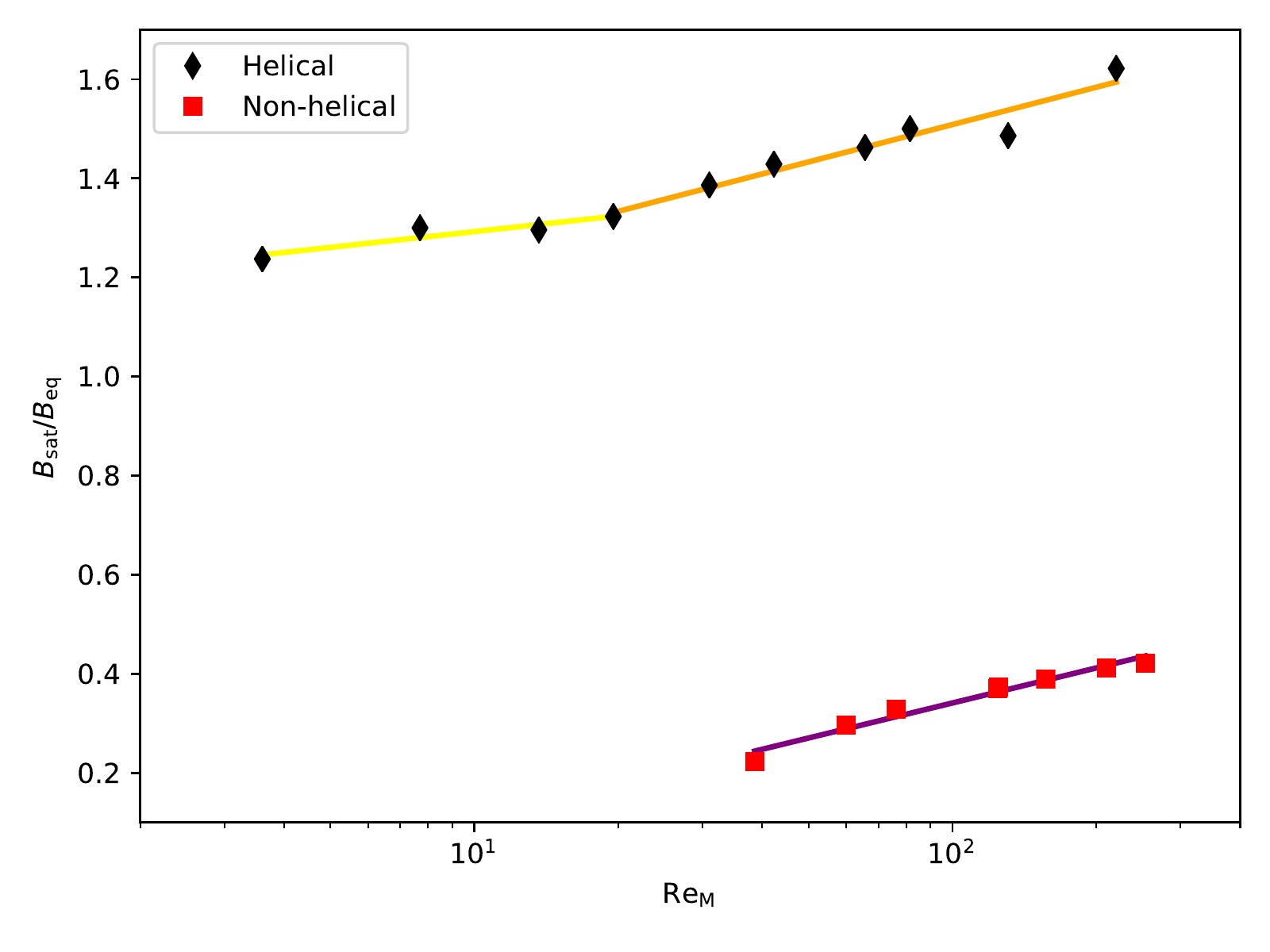}
\caption{Estimated saturation magnetic field $B_\mathrm{sat}$ normalized to
$B_\mathrm{eq}$.  The solid lines represent logarithmic fits. We have included
those datasets with $256^3$ and $512^3$ resolution from which a valid estimate
could be obtained. \label{fig:saturation} }
\end{figure}

Our main focus is on examining the
kinematic growth stage of the SSD, with (helical forcing) and without (non-helical forcing) a co-existing LSD.
Some conclusions about the saturated stage can also be drawn, but
for its complete study, many of the helical runs would need to be continued
longer, as the saturation of the LSD is known to occur on a resistive time scale only \citep[e.g.][]{Brandenburg2001}.
Unfortunately, for the highest
$\Rem$, our datasets are not long enough to determine
their saturation field strength. 
To compensate this, we performed a prolonged run with $\nu = \eta = 1.5\dee{-4}$, $\Rem = 210$,
extending to one diffusion time based on the forcing scale. This was the highest $\Rem$ allowing numerical stability towards saturation. However, otherwise we have to restrict our analysis on low and intermediate $\Rem$.
The time development of the large-scale fields
can be fitted well with a function
$B_\mathrm{sat}\tanh{(t/d_0+d_1)}$, where $d_0$ and $d_1$ are 
fitting parameters,
and we use it to 
determine the saturation magnetic field $B_\mathrm{sat}$.
Restricting to the helical runs with intermediate $\Rem$,
we find that $B_\mathrm{sat}/B_\mathrm{eq}$ 
increases roughly
logarithmically as a function of $\Rem$ (See Figure \ref{fig:saturation}).

The saturation values for the non-helical cases are easier to 
determine as they saturate quickly after their exponential growth, and compared to the helical cases, the field strengths are roughly by a factor of six smaller.  
Their $B_\mathrm{sat}/B_\mathrm{eq}$ grows also in with $\Rem$. 

For helical forcing, in both the cases with only an LSD ($\Rem<40$) and a combined LSD-SSD, $B_\mathrm{sat}/B_\mathrm{eq}$ as a function of $\Rem$ obeys a logarithmic law as shown in Figure \ref{fig:saturation}, albeit with different slopes. For combined LSD-SSD, the slope is roughly the same as in 
the non-helical cases. Therefore we hypothezise that the emergence of the SSD is to the disadvantage of the LSD, most likely due to a reduction of $\alpha$, and thus prevents its saturation strength from further growing with $\Rem$.
Consequently, the observed growth of $B_\mathrm{sat}/B_\mathrm{eq}$ with $\Rem$ would be exclusively due to the SSD.

\subsection{Power spectra}\label{sec:spectra}

One substantial difference between SSD and LSD consists in the scale distribution of the magnetic energy during growth and saturation. To investigate it, we have calculated magnetic power spectra $E_B(k)$ (with normalization $\int E_B(k) dk = \int \mathbf{B}^2 dV/2\\mu_0$  
for individual simulation snapshots.

\begin{figure}[htb!]
\plotone{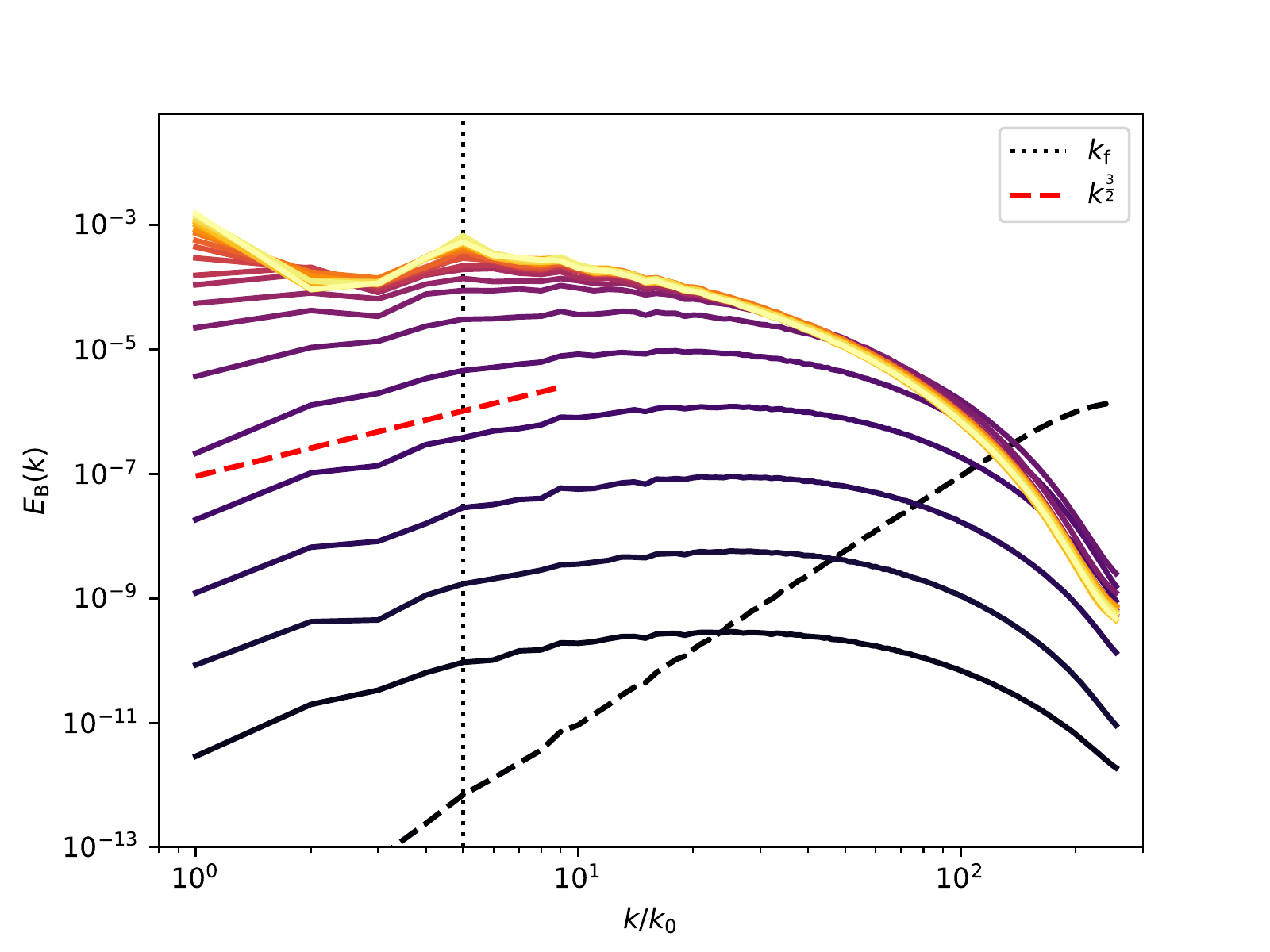}
\plotone{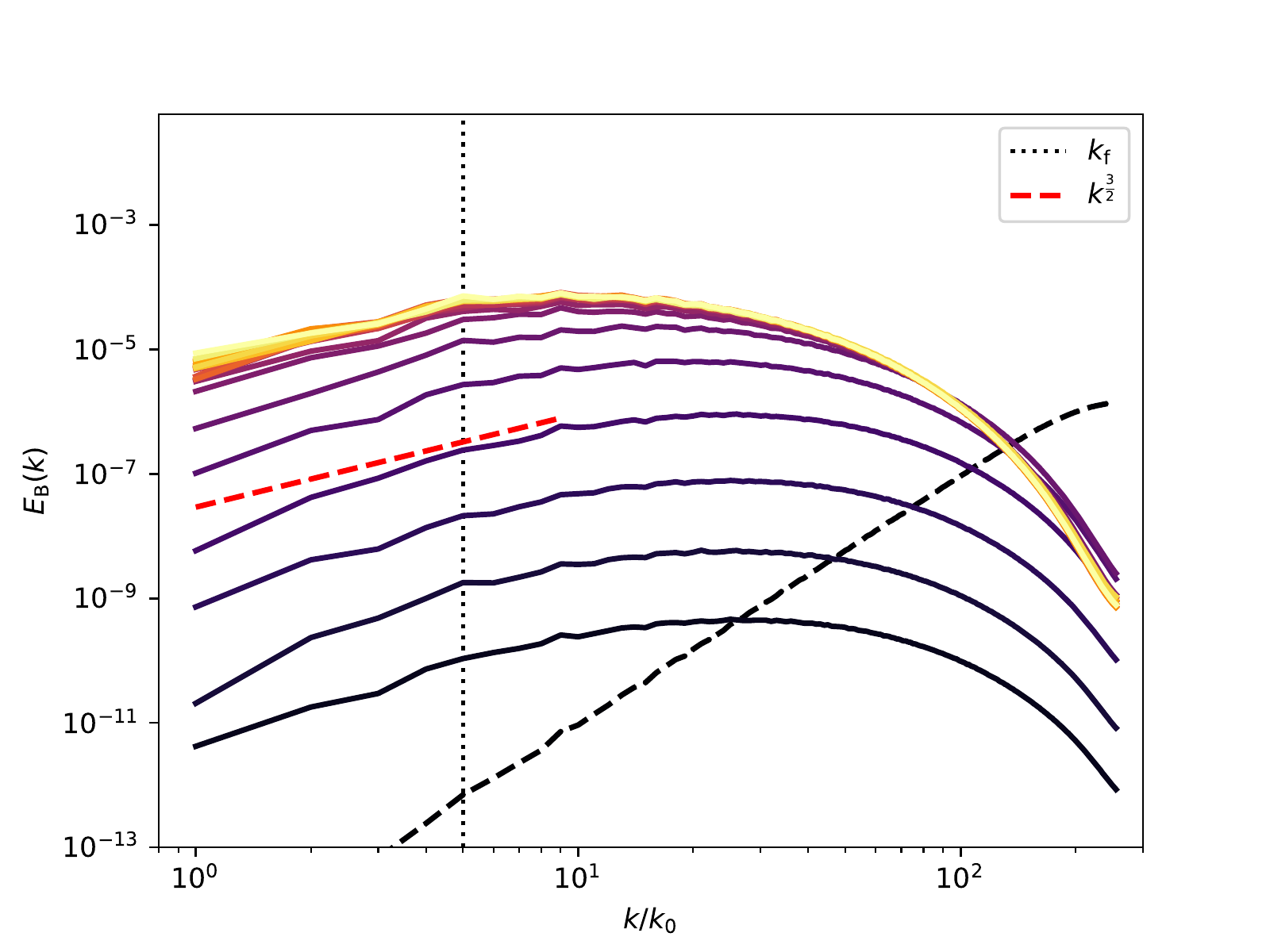}
\caption{Time-dependent powerspectra of magnetic energy, $E_B(k)$, for $\eta = 1.25\dee{-4}$ with resolution $512^3$ for helical (top) and non-helical (bottom) forcing
and $0\le t \le600$.
Both show Kazantsev scaling $\propto k^{3/2}$ during growth (red dashed). Time difference between spectra $\Delta t=30$.
For $t=0$, the spectrum reflects the initial random (Gaussian)  field, hence $\propto k^2$.
The dashed curves correspond to $t=0$.
\label{fig:pspec_a} 
}
\end{figure}

\begin{figure}[thb!]
\plotone{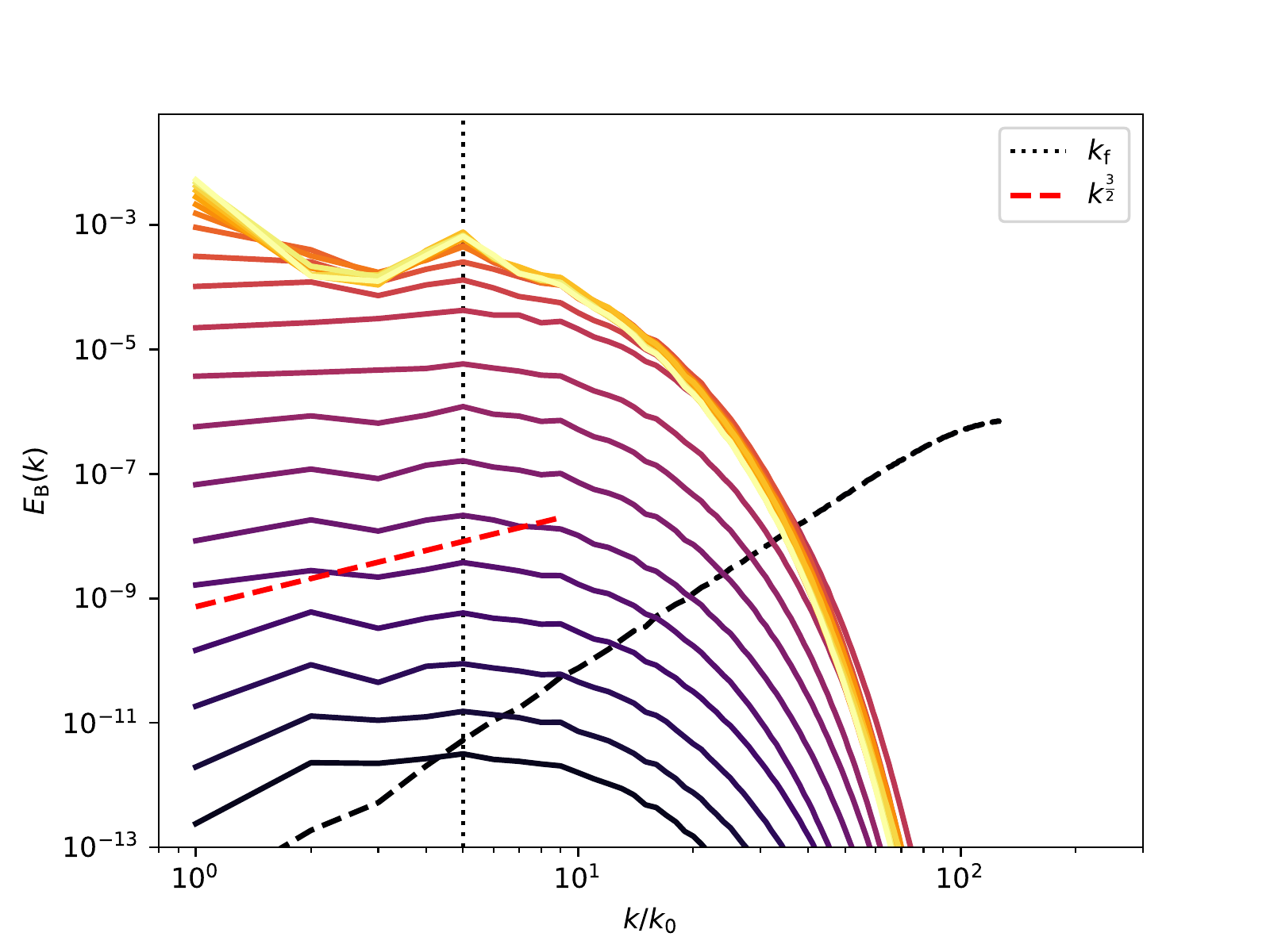}
\caption{Time-dependent powerspectra of magnetic energy, $E_B(k)$, for $\eta = 2\dee{-3}$ with resolution $256^3$ for helical forcing for $0\le t \le600$.  Time difference between spectra $\Delta t=30$.
\label{fig:pspec_flat} 
}
\end{figure}

Figures \ref{fig:pspec_a} and \ref{fig:pspec_flat} show time-dependent spectra $E_B(k;t)$ for the highest studied $\Rem$ and a low one, respectively.
In the case of helical forcing (LSD) with high $\Rem>\Remcr$ (Figure \ref{fig:pspec_a}, top), we witness a persistent peak at the forcing wavenumber $\kf$ and at late times gradually growing power for $k\gtrsim 1$, i.e. in the large-scale field. At late times, there is a typical forward energy cascade towards small scales $k>\kf$,. 
For SSD, (Figure \ref{fig:pspec_a}, bottom), large scales follow persistently the Kazantsev scaling $\propto k^{3/2}$ \citep{Kazantsev1968}, while the spectrum peaks above $\kf$ at $k = 9$; beyond that a similar forward cascade exists as in LSD. 
Qualitatively, this picture is the same for all $\Rem>\Remcr$.

For high $\Rem$, as shown in Figure \ref{fig:pspec_a}, even with helical forcing we can clearly see SSD-type (Kazantsev) spectra during the growth stage, which signifies the presence of SSD during the growth. The SSD-type spectra can also appear during the growth of medium $\Rem$ cases, which are not too diffusive. 
However, at low $\Rem$, SSD and LSD spectral shapes are clearly different during this stage: 
In helically driven systems, the spectrum is essentially flat at large scales until the emergence of the $k=1$ mode, see Figure \ref{fig:pspec_flat}. 
This resembles the spectral growth of \citet{Brandenburg2001} with a flattened curve during the growth stage and the eventual emergence of $k=1$ mode (Their Figures 2 ans 3 respectively). 

\begin{figure}[h!]
\plotone{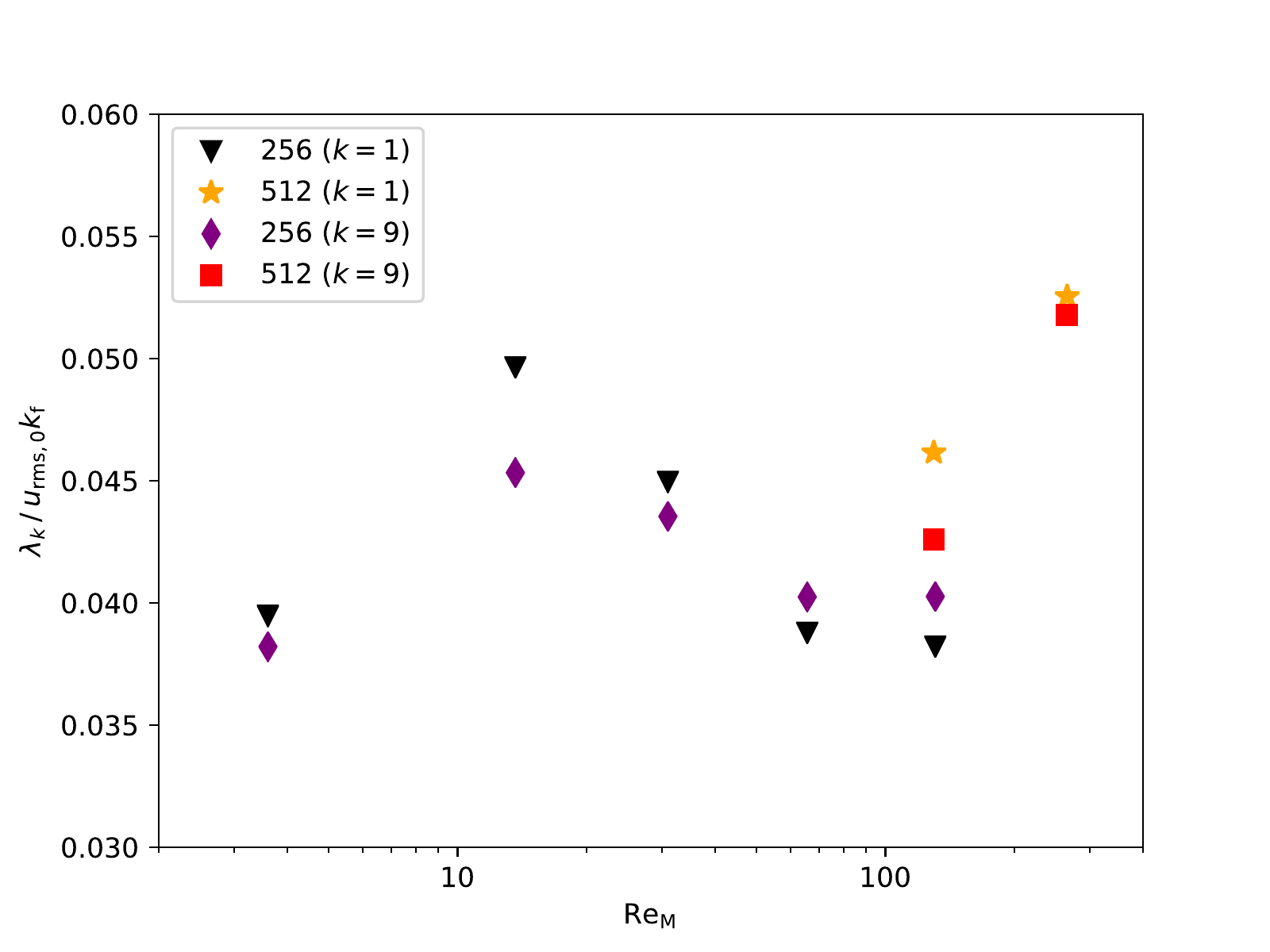}
\plotone{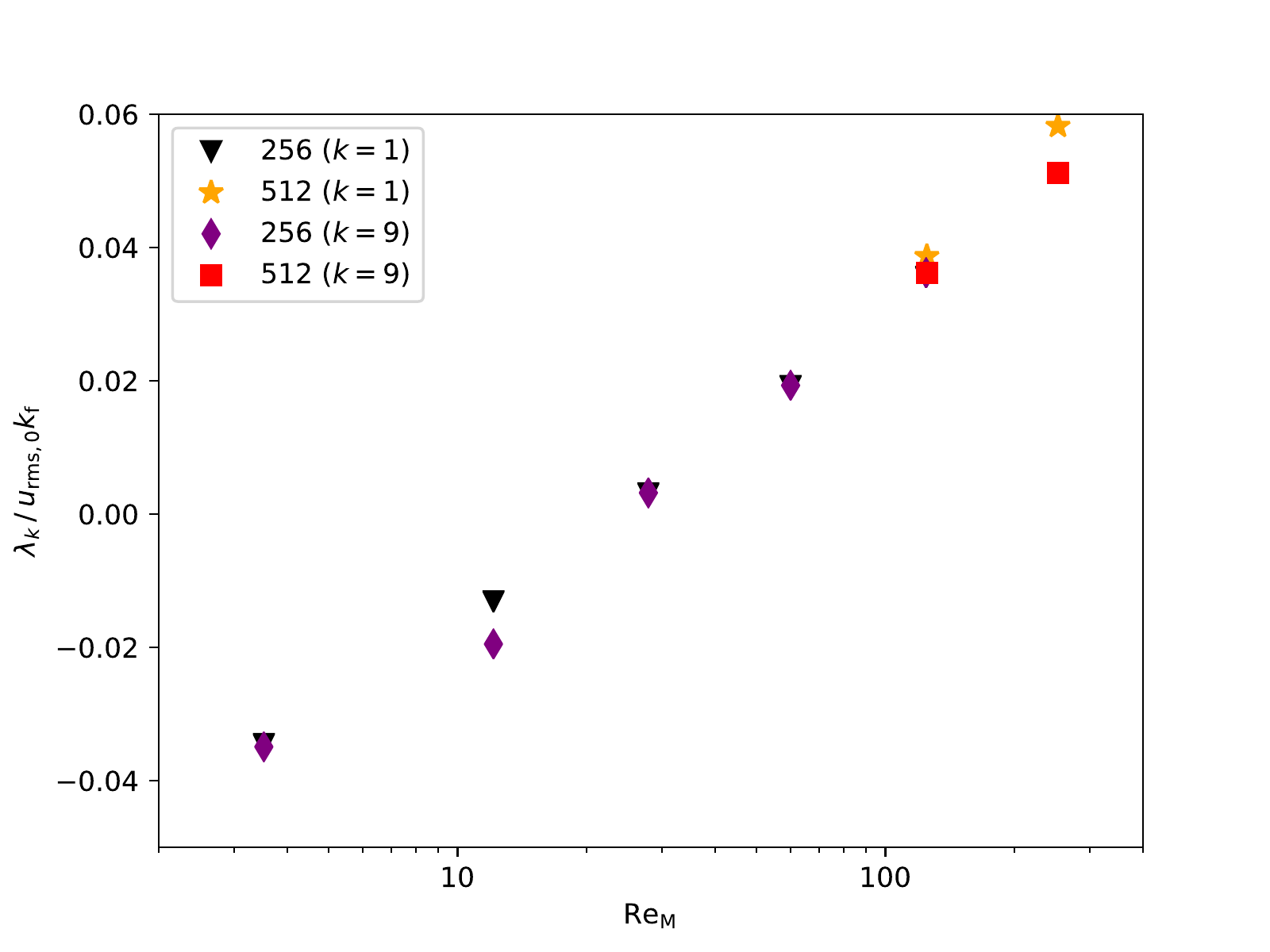}
\plotone{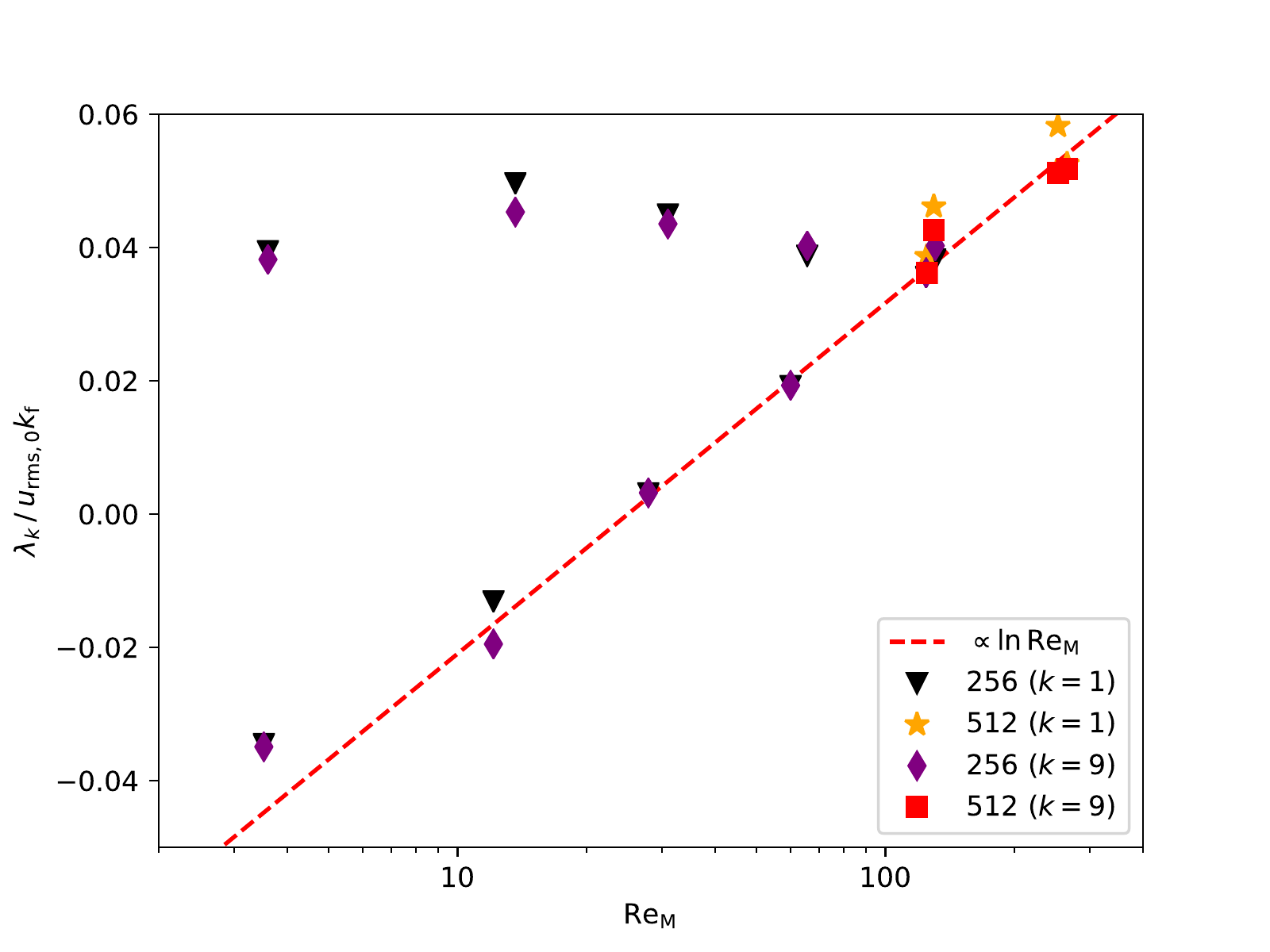}
\caption{Growth rates $\lambda_k$ of the power spectra channels $k=1$ and $k=9$ for helical (top) and non-helical (middle) forcing. Bottom: combination of top and middle panel. \label{fig:growthrate_channel} }
\end{figure}

As helical and non-helical growth rates converge at high $\Rem$ (see Sec. \ref{sec:growth} and Figures \ref{fig:growthrate} and \ref{fig:growthrate_nohel}),
we may explain this behavior by the assumption that in this range the SSD growth rate is higher than the LSD one, thus the former is dominating the latter during growth. But if SSD saturates earlier than LSD and also at lower magnitude,  the spectrum has to undergo the observed change in its shape with finally dominating large scales.   
For medium $\Rem$ the situation would be less clear.
The Kazantsev scaling is kept during most of the growth stage, 
but the growth of the magnetic energy is more rapid than for the respective non-helical runs.
Therefore helical turbulence can enhance the accumulation of magnetic energy without affecting the spectral shape during the initial growth.

In an attempt to separate SSD and LSD behaviour, we have estimated the growth
rates as functions of wavenumber. The resulting $\lambda_k$, displayed in
Figure \ref{fig:growthrate_channel} for $k=1$ and $k=9$, are 
in general proportional to those estimated from $B_{\mathrm{rms}}$. 
Again, the growth rates of SSD and LSD merge at high $\Rem$ and the SSD ones follow the logarithmic scaling. 
For both helical and non-helical forcing, both of the referred scales ($k=1$ and $k=9$)
grow at similar rates, 
with a mild tendency of 
$\lambda_1$ being marginally higher than $\lambda_9$.  In addition, the helical runs seem to show higher growth rates at high $\Rem$ than the nonhelical runs. 
Yet, because of fitting uncertainties, all this should be taken with caution.

\begin{figure}[tb!]
\plotone{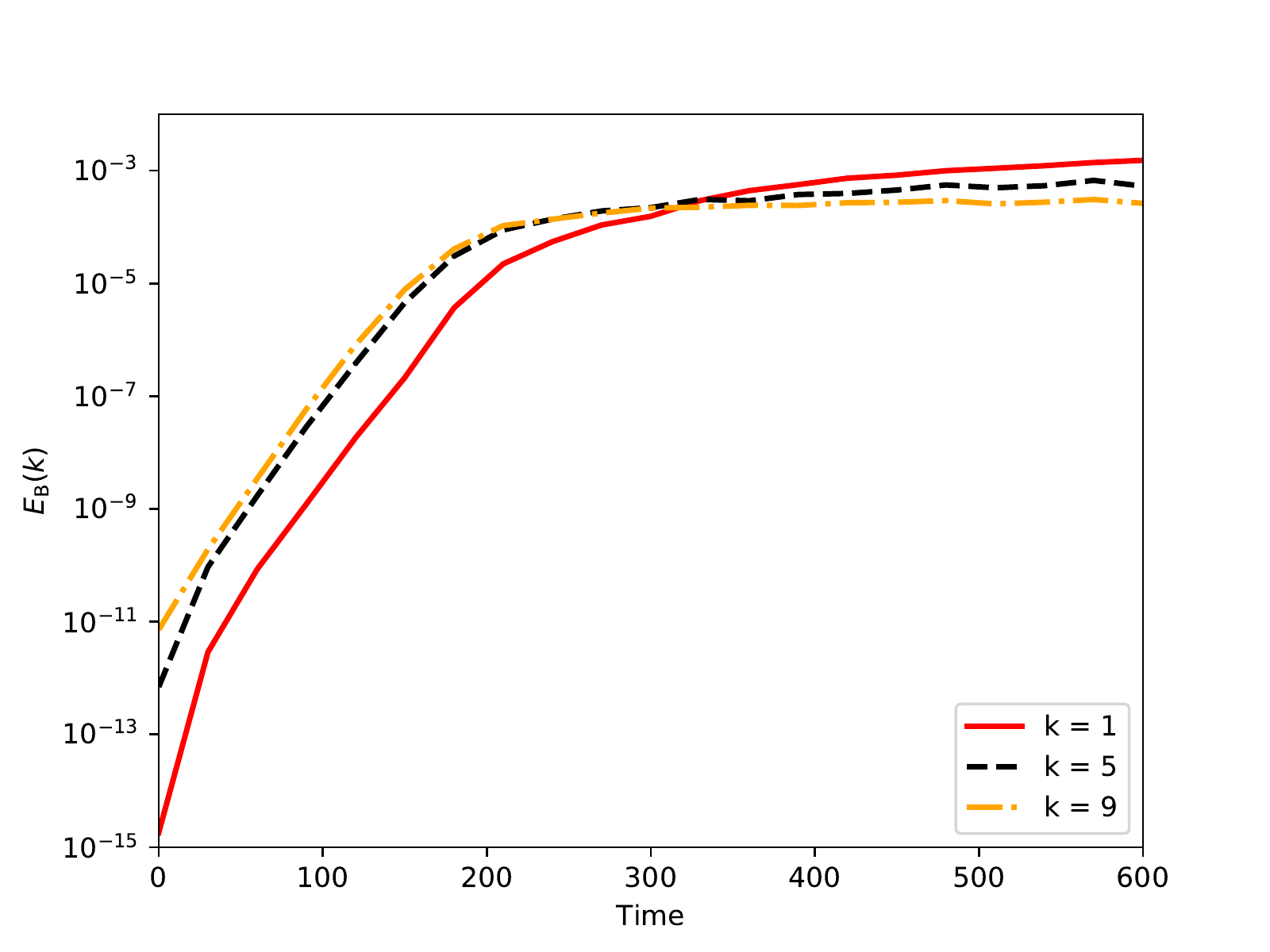}
\plotone{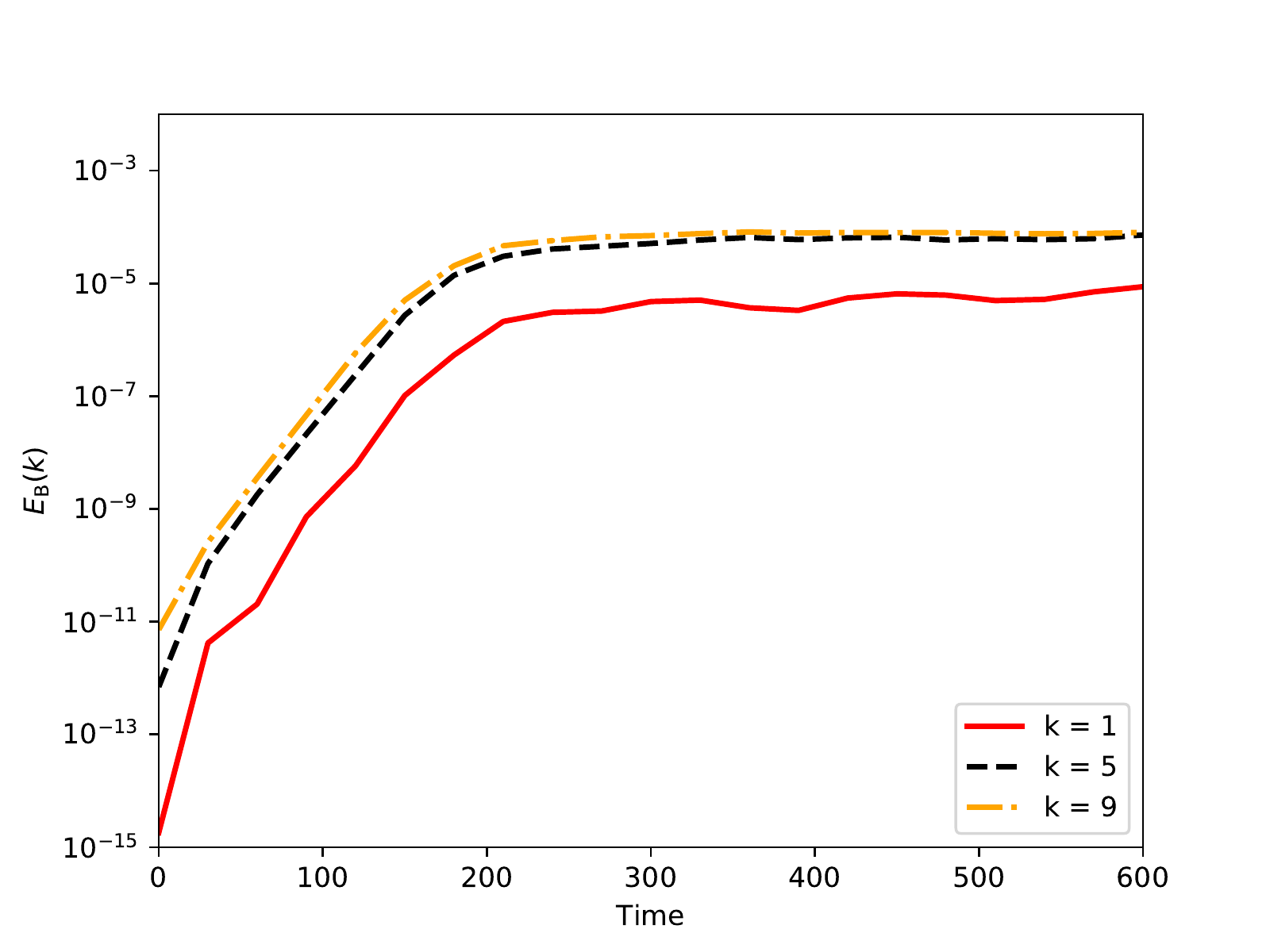}
\caption{Time development of selected  spectral channels for $\eta = 1.25\dee{-4}$ with resolution $512^3$ for helical (top) and non-helical (bottom) forcing. \label{fig:psts_a} 
}
\end{figure}

\begin{figure}[htb!]
\plotone{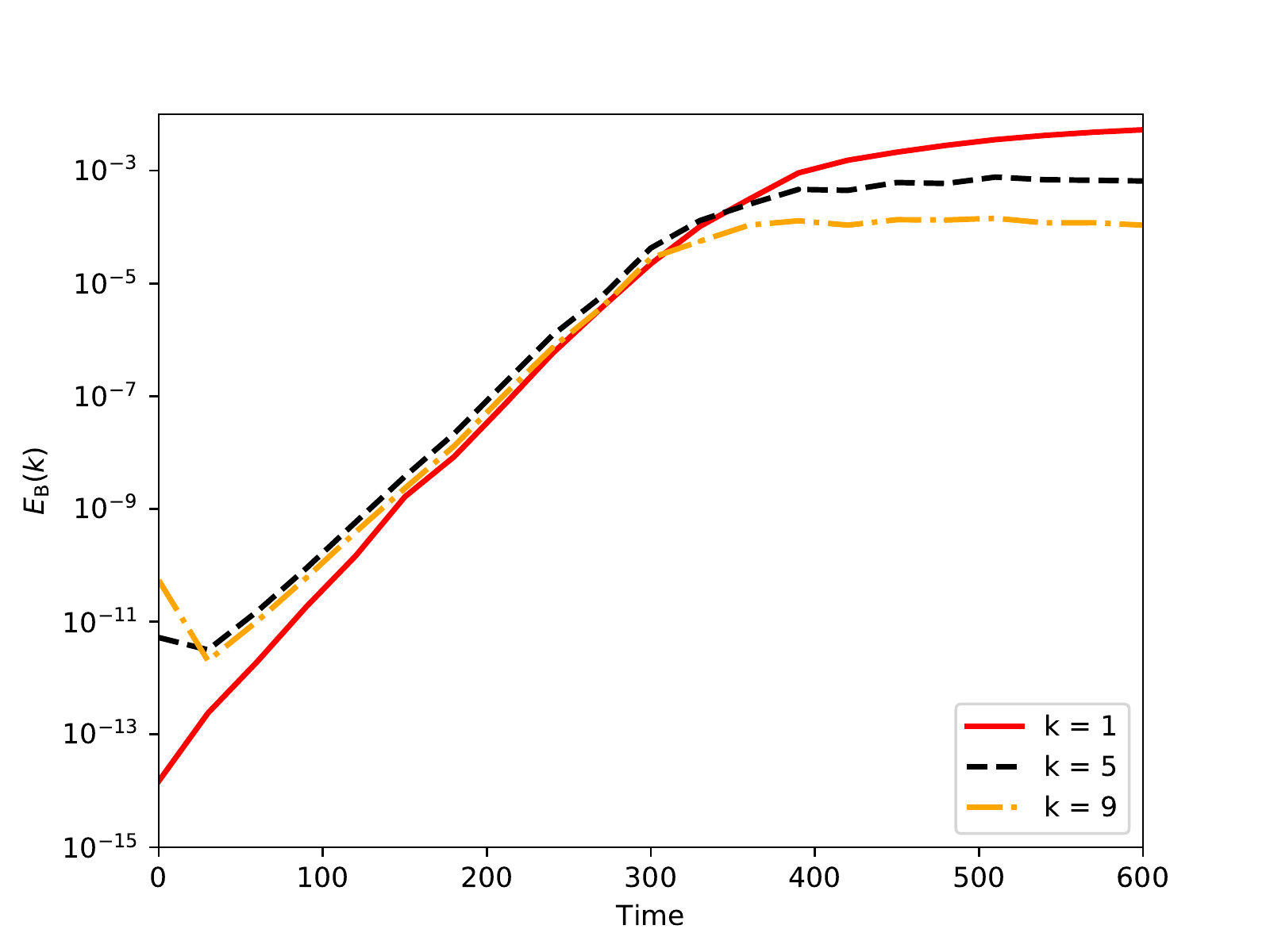}
\caption{Time development of selected spectral channels for $\eta = 2\dee{-3}$ with resolution $256^3$ for helical forcing.  \label{fig:psts_flat} 
}
\end{figure}

Figure \ref{fig:psts_a} displays the corresponding time development of the spectral channels $k=1,5,9$ for the high $\Rem$ helical and non-helical runs featured in Fig. \ref{fig:pspec_a}. In both cases, the exponential growth occurs at similar rates for all three channels
with the $k=5,9$ ones dominating. The only difference is that when LSD is present, the $k=1$ channel continues to grow past the exponential growth stage,
while the other channels are almost saturated, to become finally dominating.
At low $\Rem$, with LSD alone (Figure \ref{fig:psts_flat}), all three channels have approximately equal magnitude with minor deviations in the beginning. After the exponential growth phase the spectral channels differ with $k=1$ becoming the strongest, 
as it happens in the spectra of Figure \ref{fig:pspec_flat}. 

\begin{figure}[htb!]
\plotone{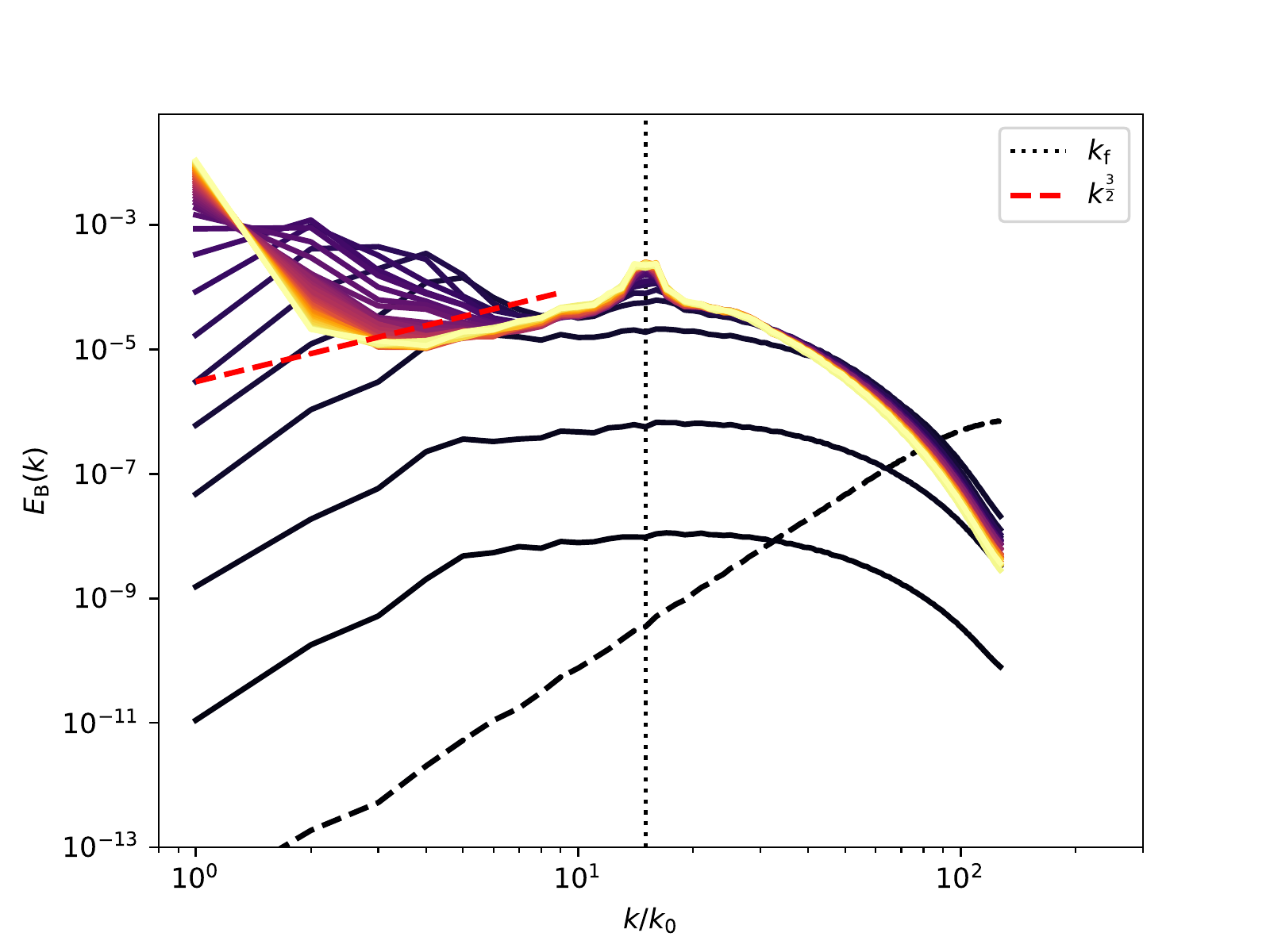}
\caption{Time-dependent power spectra for $\eta = 2.5\dee{-4}$ with resolution $256^3$ for helical forcing with $\kf = 15$.  Time difference between spectra $\Delta t=100$. \label{fig:pspec_k15} 
}
\end{figure}

\begin{figure}[htb!]
\plotone{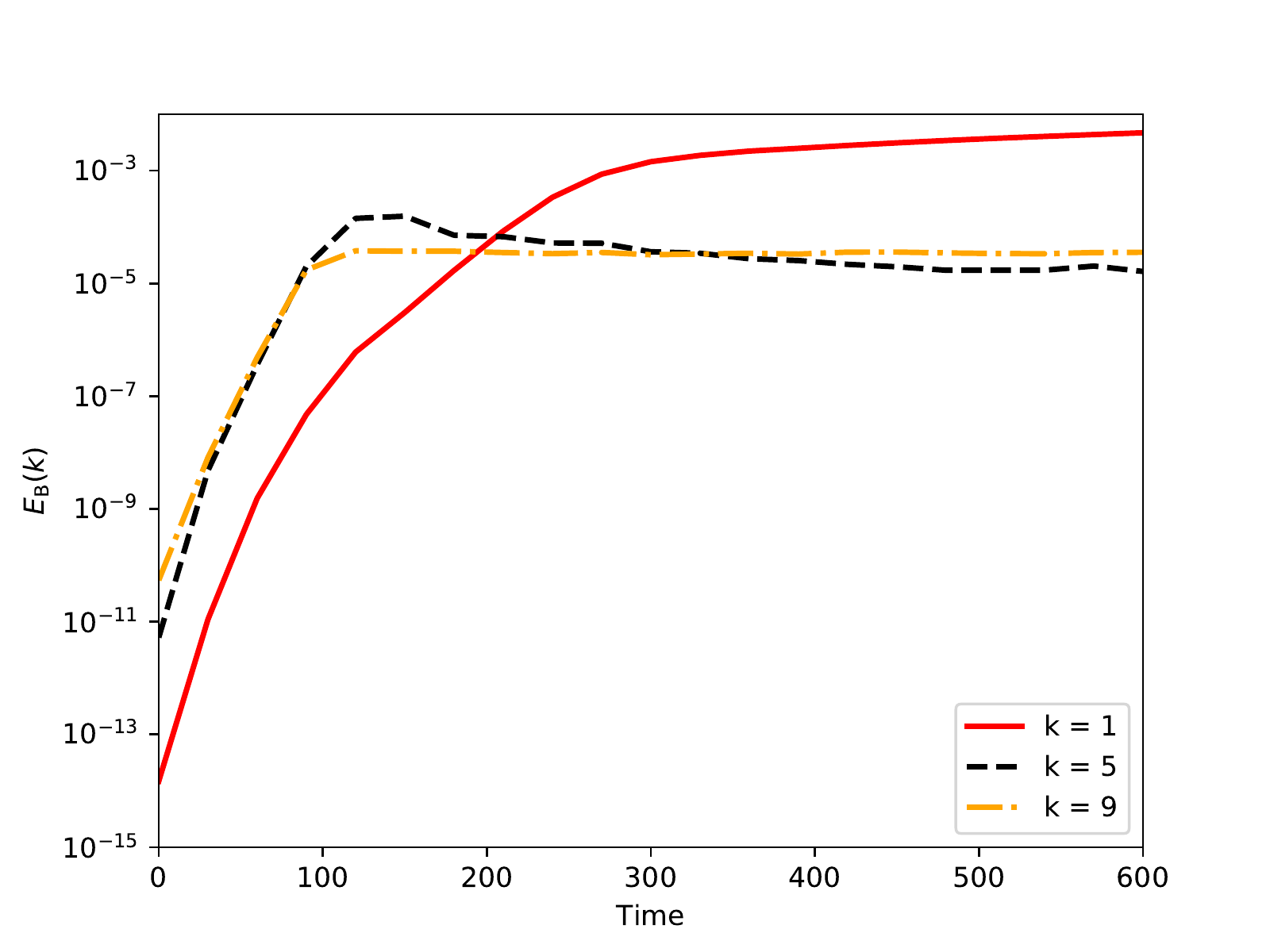}
\caption{As Figure \ref{fig:pspec_k15}, but time evolution of the spectral channels $k=1,5,9$. \label{fig:psts_k15} 
}
\end{figure}

For the sake of testing and comparison, we changed the forcing scale to $\kf= 15$ to separate it more safely from the largest scale of the emergent mean field. During growth we see a more gradual buildup of the
inverse energy cascade towards  large scales, see Figure \ref{fig:pspec_k15}:  
the energy peak of the growing large-scale magnetic field moves gradually towards larger scales until reaching its largest values at $k = 1$. This is similar to the same phenomena visible in Figure 7 of \citet{Brandenburg2012review}. 
The time development of the spectral channels $k=1,5,9$ is shown in Figure \ref{fig:psts_k15}, indicating a clearly different growth rate of the $k = 1$ channel.
The $k = 1$ channel curve consists of two different exponentials at concurrent stages. However, the data is too sparse in that range to produce meaningful fit estimates. 
There is even decay visible 
in the $k=5$ channel which is explainable with the magnetic energy inverse-cascading onto larger scale over time as visible in Figure \ref{fig:pspec_k15}.

\subsection{Mean-field analysis}\label{sec:soca}

For LSD, mean-field theory provides some testable predictions: First, an estimate of the growth rate based on turbulent transport coefficients. Second, quenching of the $\alpha$-effect as a function of $\Rem$. 

Unfortunately, at this moment \textit{Astaroth} is not capable of handling the test-field method \citep{Schrinetal07} or other method 
for measuring the turbulent transport coefficents
at runtime. 
Therefore, we calculated estimates for the coefficients of $\alpha$-effect and turbulent diffusion, $\etat$,  using results from the second-order correlation approximation (SOCA) and other closures
which have proven to be surprisingly useful \citep{Sur2008, Vaisala2014}. 

According to the mean-field approximation, the growth rate of the mean field $\meanBB$ for isotropic stationary turbulence, hence constant $\alpha$ and $\etat$, is 

\begin{equation}\label{eq:MFgrowth}
    \lambda = |\alpha| k - (\etat + \eta) k^2 
\end{equation}
where $k$ is the wavenumber of the mean field and $\alpha = \alpha_K + \alpha_M$. In the limit of ideal MHD, the constituents of $\alpha$ are related  to kinetic and current helicity, respectively, by
\begin{equation}\label{eq:alphaFOSA}
    \alpha_K = -\frac{1}{3} \tau \left\langle \boldsymbol{\omega} \cdot \mathbf{u} \right\rangle \quad \mathrm{and} \quad
    \alpha_M = \frac{1}{3} \tau \langle \mathbf{j}'\cdot \mathbf{b}' \rangle /\mu_0 \rho_0,
\end{equation}
 with correlation time $\tau$, and vorticity $\boldsymbol\omega=\nabla\times \mathbf{u}$ and primes indicating the fluctuating parts. 
Note two possible interpretations for $\alpha_M$:
First, it reflects the contribution of a magnetic background turbulence, like that provided by an SSD, to $\alpha$. Here, $\alpha_M$ can be obtained already by SOCA \citep{RaeRhei07}. Second, it can be interpreted as reflecting the quenching of $\alpha$ by $\meanBB$ such that with its magnitude growing, $\alpha_M$ also grows, but opposite in sign to $\alpha_K$, resulting in a reduced total $\alpha$. This can be obtained via closure approaches like the $\tau$
 or eddy-damped quasi-normal Markovian approximations \citep{Pouquet1976}.
Turbulent diffusivity in incompressible flows is estimated as 
\begin{equation}\label{eq:etaFOSA}
    \eta_\mathrm{t} = \frac{1}{3} \tau \langle {\mathbf{u}}^2 \rangle.
\end{equation}
If the Strouhal number $u_{\mathrm rms}\tau/\ell$, $\ell$ a characteristic scale of the flow, is assumed to be unity, we can estimate  $\tau = 1/\kf u_\mathrm{rms}$. To obtain the fluctuating fields $\mathbf{b}'$, $\mathbf{j}'$ we have removed the large scale field via filtering out 
the contributions of the $k = 1$ mode from the magnetic field snapshots.

\begin{figure}[th!]
\plotone{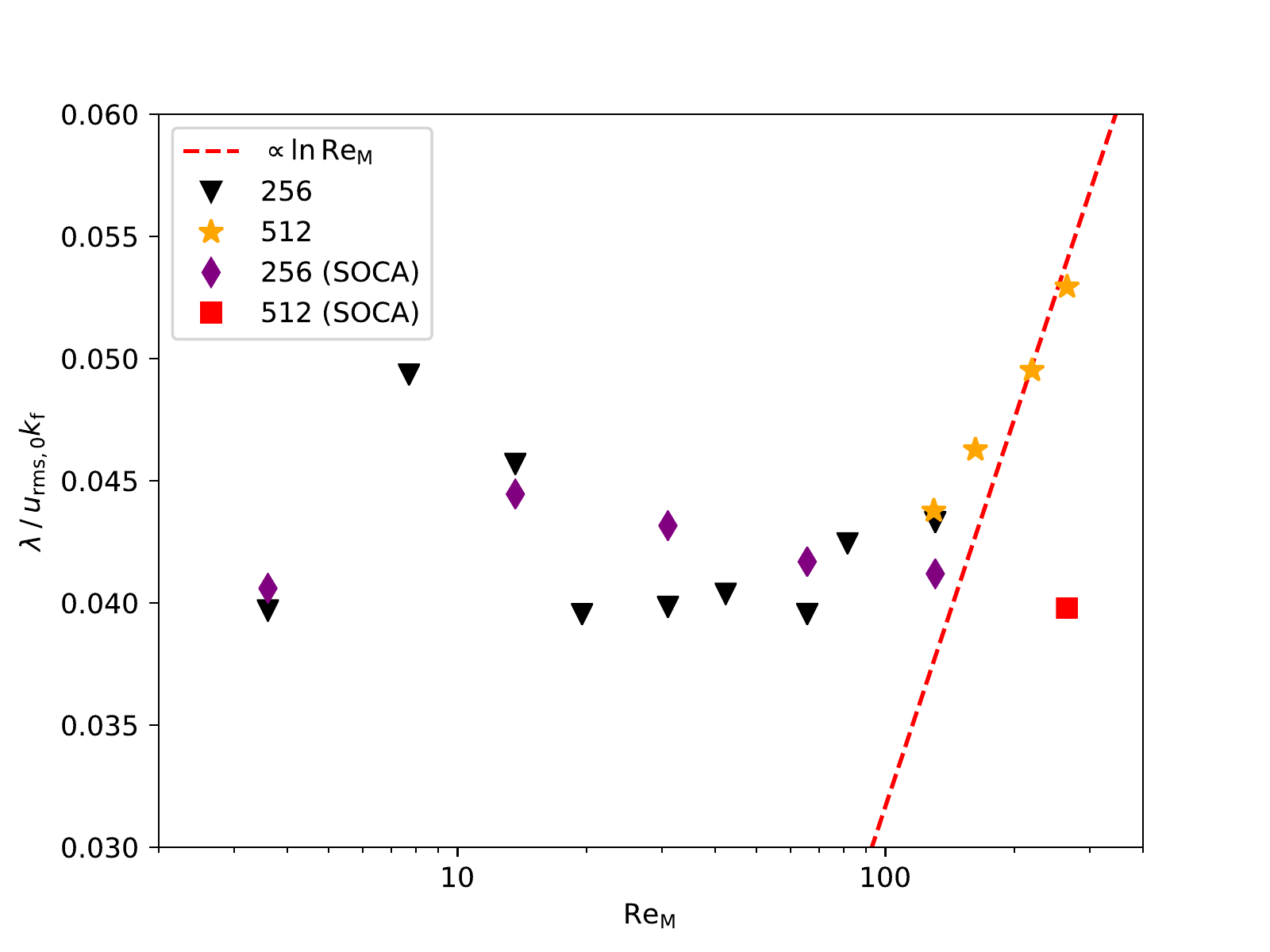}
\caption{Growth rates \eqref{eq:MFgrowth} based on closure estimates of $\alpha$ and $\eta_\mathrm{t}$ at the exponential growth stage, compared to directly estimated ones.  \label{fig:socagrowth} }
\end{figure}

We estimated the growth rates by first calculating $\alpha$ and $\eta_\mathrm{t}$ for individual snapshots using data from runs with high snapshot frequency during exponential growth. Then we used Equation (\ref{eq:MFgrowth}) to get $\lambda$ for an individual snapshot, and subsequently time-averaged over the growth phase. 
We find results which are at least approximately aligned with the directly measured values 
with better agreement at low $\Rem$,
see Fig. \ref{fig:socagrowth}. Towards high $\Rem$, the results begin to diverge, most strongly at the highest $\Rem$. 
This appears to indicate that the closure estimates can work surprisingly well, but do not catch all of the details.

But how does $\alpha$ behave at the saturation stage?  \citet{Brandenburg2008} measured $\alpha$ and $\etat$ quenching, with both decreasing as functions of $\Rem$.

\begin{figure}[ht!]
\plotone{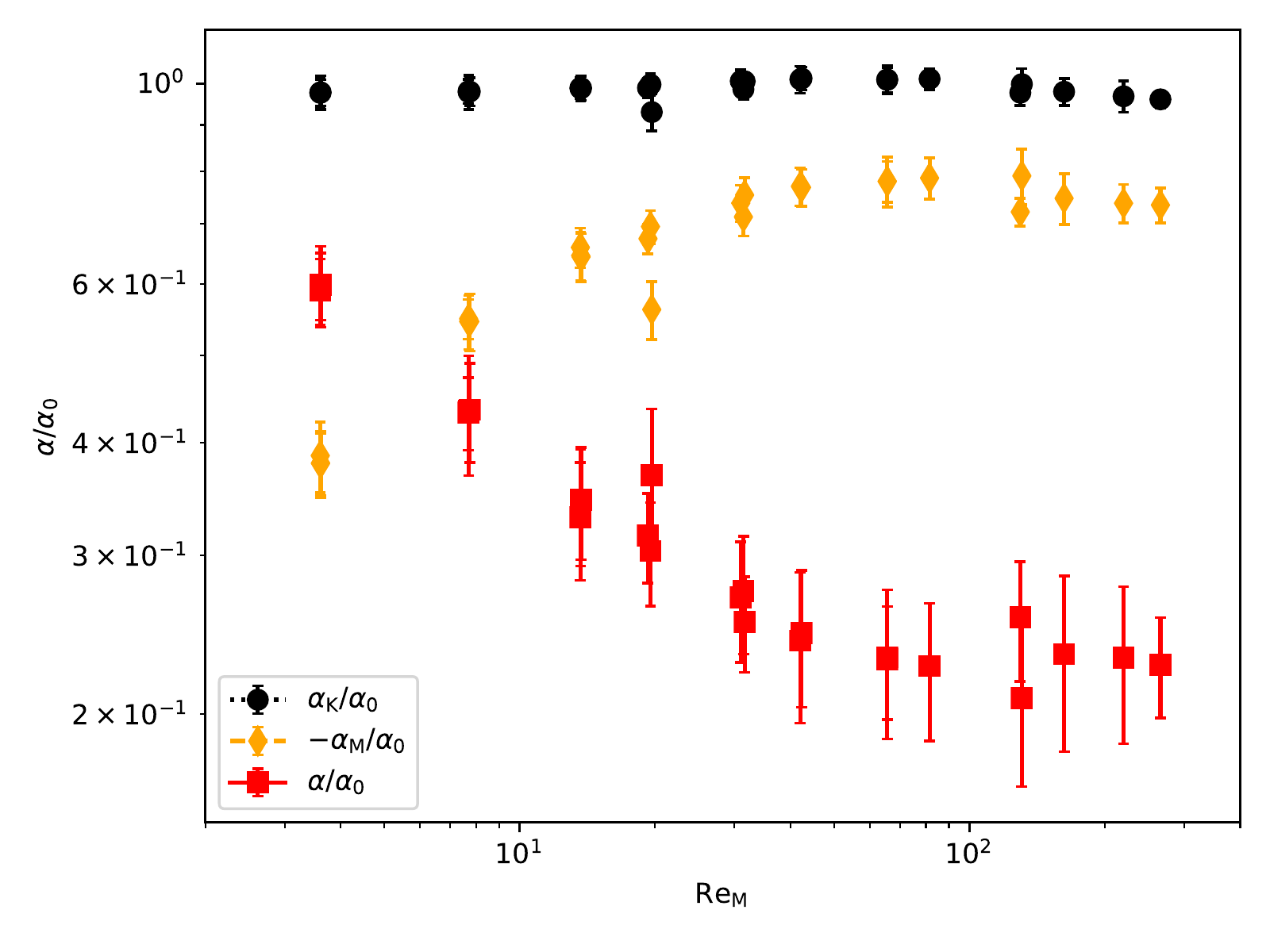}
\caption{$\alpha_{\rm K}$ and $\alpha_{\rm M}$ at the saturated stage as functions of $\Rem$, normalized by $\alpha_0 =  u_\mathrm{rms}/3$. 
Bars represent the fluctuation level of $\alpha$ derived from their standard deviations. 
The jump at $\Rem \sim 20$ is due to an unclear disagreement between resolutions $128^3$, $64^3$ and $256^3$, that at $\Rem \sim 100$  due to different saturation stage lengths in the runs with resolutions $256^3$ and $512^3$ . 
\label{fig:alphaquench} }
\end{figure}

For comparison, 
we calculated $\alpha_{\rm K,M}$ from the saturation stages of our runs and discovered similar results, see Figure \ref{fig:alphaquench}. The normalized $\alpha_K$ approaches a constant whereas $\alpha_M$ keeps getting stronger with $\Rem$.
Our results are close to \citet{Brandenburg2008} (their Fig. 3) within our range of $\Rem$, including  substantially more points though. The total $\alpha$ decreases with growing $\Rem$ reaching eventually a tentative plateau. 

\subsection{Magnitude distribution of the magnetic field}

\begin{figure*}[ht!]
\plotone{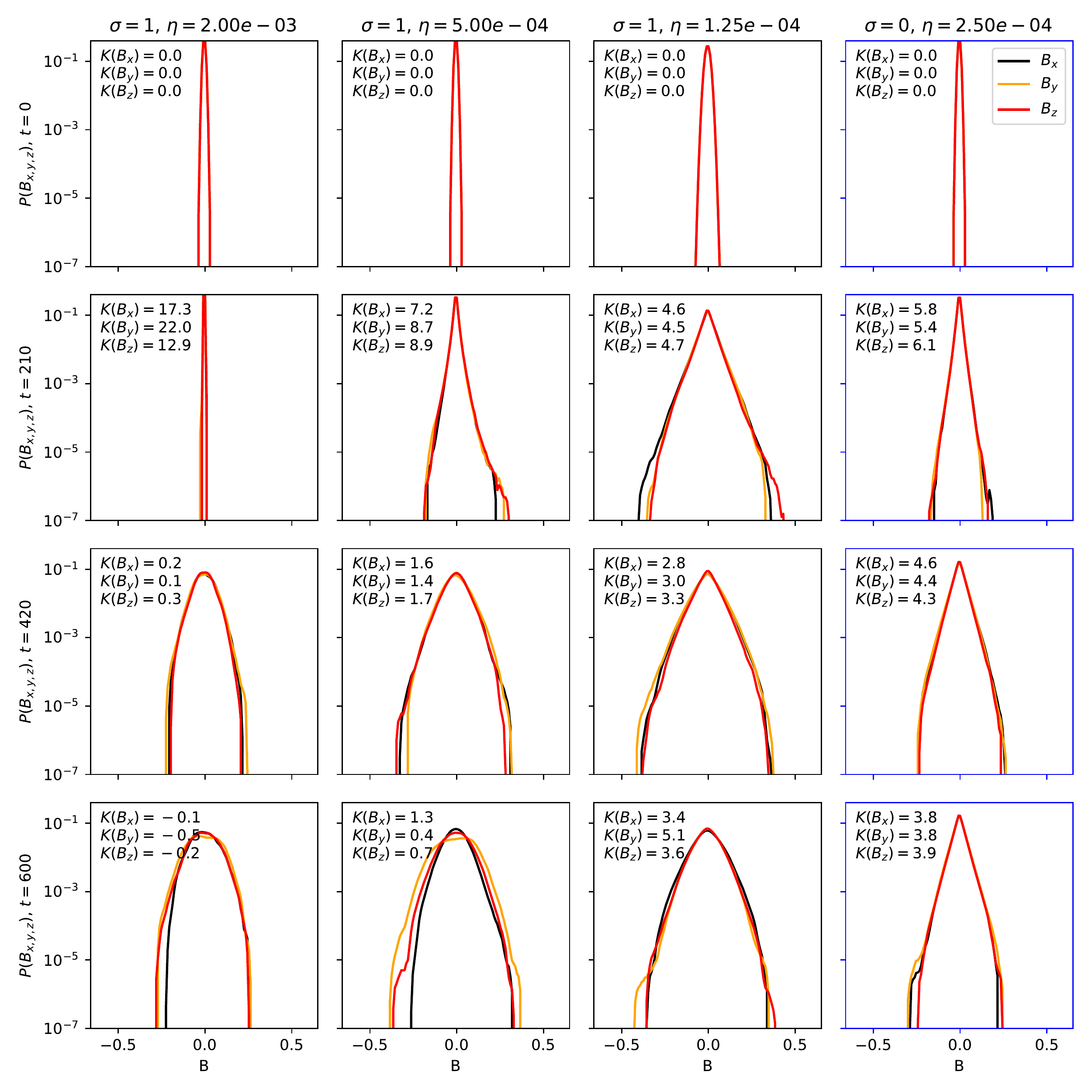}
\caption{Probability density functions of $B_{x,y,z}$ for non-helical forcing (first column) and for helical forcing with low, intermediate and high $\Rem$ (second to fourth column), as a function of time (rows). \label{fig:BPDF} 
$K(B_x)$, $K(B_y)$ and $K(B_z)$ denote the respective kurtoses using Fisher’s definition.}
\end{figure*}

Next we investigate how the probability density functions (PDFs) of the
magnetic field differ in the regimes where SSD or LSD alone and SSD and LSD
together are acting. While spectra gave us information about the distribution
of the magnetic field across different scales, PDFs can further reveal details
of the structure of the field in different spatial directions. Additionally, we
calculated kurtoses of these distributions using the
Fisher’s definition, for which the kurtosis of a normal distribution is 0. 
 
Figure \ref{fig:BPDF} shows the temporal evolution of the PDFs of all three components of the dynamo-generated $\mathbf{B}$ for non-helical forcing and helical forcing with three different $\Rem$.
Two basic types of PDFs are obtained:
A SSD produces a symmetric exponential distribution, which expands 
over time, but keeps otherwise the same shape
and eventually stops changing at the saturation stage, with its kurtosis ranging from $\sim 4$ to $\sim 6$.  
An LSD produces a more Gaussian--shaped distribution. 
As a pure Beltrami field has a PDF of top-hat shape,
this can result in a PDF of the total field with a both widened and flattened peak if $\meanBB$ is strong enough.
In Figure \ref{fig:BPDF}, such an indication of a top-hat profile is best
visible in the PDFs of $B_y$ at saturation for helically forced cases with
$\eta = 0.002$ and $0.0005$, with their kurtoses approaching 0, whether or not
the top is flattened. This is not surprising because the top hat effect happens
close to the peak and the PDF has still significant tails. 

The widened Gaussian PDF profile of LSD appears most pronouncedly at (or just before) the saturation stage, whereas during the exponential growth stage, the LSD cases show at high and intermediate $\Rem$ rather an SSD type of PDF, including similar values of the kurtosis.
Even after initial growth, there is a tendency of the LSD-type PDFs to develop a sharp tip, reminiscent of the SSD type, because the emergence of fully saturated large scale Beltrami field takes time. 
These observations support the finding of Section \ref{sec:spectra}, that at high $\Rem$ exponential growth is seemingly dominated by the SSD, present simultaneously with the LSD. 
However, even with high $\Rem$, where small-scale fluctuations are strong, the LSD will turn the field profile into a Gaussian type over time.  In Figure \ref{fig:BPDF}, the highest $\Rem$ simulation is not depicted at its most saturated state, but the $k=1$ mode will keep growing mere over time and its feature will soften.

\section{Discussion}\label{sec:discussion}

We examined emergence and growth of both large and small scale dynamos and found that with helical forcing, approaching  high $\Rem$, both SSD and LSD become clearly simultaneous phenomena. This is visible both from the time evolution of the  powerspectra and from the probability density functions of $\mathbf{B}$. For high $\Rem$, the growth rates of helical and non-helical cases converge, indicating the dominance of SSD in the helical ones.

\begin{figure}[ht!]
\plotone{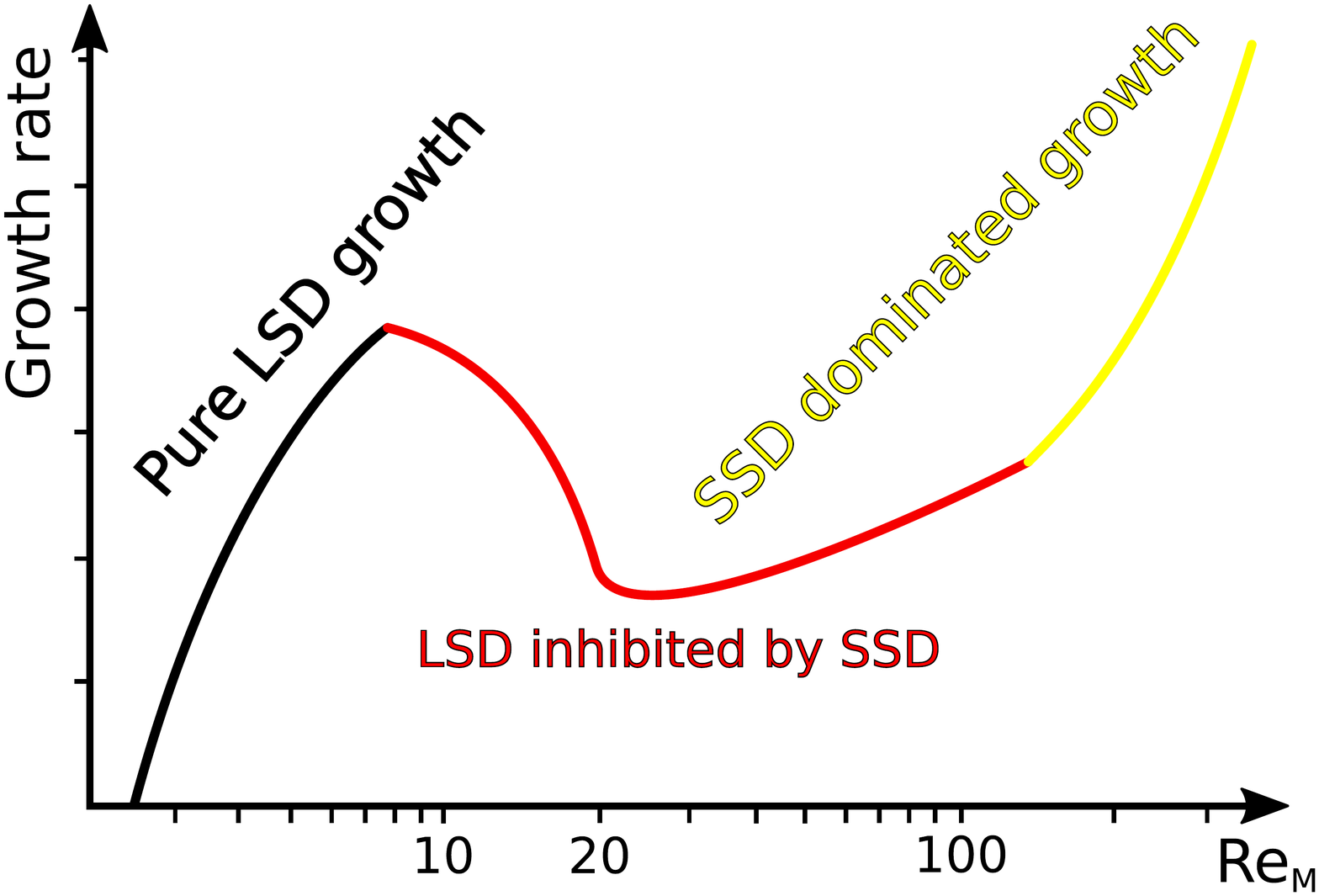}
\caption{Schematic depiction of our hypothesis for the change of growth rate as a function of $\Rem$ in helically forced simulations, cf. Figure \ref{fig:growthrate}.  \label{fig:LSDvsSSD} }
\end{figure}

Under helically driven turbulence, an exact delimitation between LSD and SSD is difficult. At medium $\Rem > \Remcr$, there is  a tentative indication of an SSD in the powerspectra during the dynamo growth. While the growth rate is dominated by the LSD, an SSD can already be operating, so that SSD and LSD are coupled at this stage.

Based on the dependency of the growth rates of the helical simulations on $\Rem$, we attempt a hypothetical explanation of the interaction of the two dynamos, see Figure \ref{fig:LSDvsSSD}: At low $\Rem$, the system is diffusive enough to prevent any presence of SSD, and the -- pure LSD -- growth rate increases as a function of $\Rem$. 
In the mid range of $\Rem\gtrsim 10$, the growth rate  decreases as a function of $\Rem$. We suggest that now SSD becomes effective and starts to inhibit the growth of the LSD by reducing $\alpha$ via, in turn, increasing $\alpha_M$. However, to prove this we had to show that the $\Remcr$ of helical turbulence is smaller than that of non-helical turbulence and that $\alpha$ is really reduced. While possible, recognizing both of these affects conclusively would require more thorough mapping of $\alpha$ than what we have available. 
At high $\Rem$, SSD dominates the  exponential growth, which is subtantiated by our results. 

We can find a point of comparison in \citet{Hotta2016}, who have 
reported on the effect of SSD on LSD  in the context of 
turbulent convection
in stars with a solar--like convective envelope.  
They state that SSD shows both inhibiting and enhancing effects to the emerging magnetic field
depending on $\Rem$. In their low $\Rem$ case, a large-scale magnetic field emerges, but in their medium $\Rem$ case, the large-scale magnetic field is suppressed, 
while emerging again in their high $\Rem$ case. 
\citet{Hotta2016} explain their medium $\Rem$ case as the suppression of LSD by SSD, whereas in their high $\Rem$ case the SSD would enhance the magnetic energy generation by LSD. 
Our hypothesis, illustrated by Figure \ref{fig:LSDvsSSD} would be congruent with their scenario. However, caution should be shown because our methods are not completely equivalent to theirs.
We have focused on exponential growth, and we cannot compare the saturated stages with equal detail. 
Our limited saturation data at low, medium and high $\Rem$ shows that the
saturated magnetic field increases towards high $\Rem$, with is a tentative
indication that SSD would exclusively contribute to the the increase of
saturated field strength, but with SSD being a disadvantage of the LSD itself.
Another caveat with respect to the \citet{Hotta2016} results is that their
model diffusivities are implicit and their diffusion schemes changes between
different simulation while ours are explicit and have an uniform scheme, which
makes a direct comparison difficult. 

When it comes to the pure SSD, our results show that the $\Rem$ dependence of the growth rate matches the logarithmic prediction of \citet{Kleeorin2012} very well. In contrast, the $\propto \sqrt{\Rem}$ relation \citep{Haugen2004} does not seem functional at low $\Rem$. The logarithmic scaling appears to be empirically valid, although it  has been derived for low magnetic $\Pm$, whereas in our simulations $\Pm$ was unity.

We also estimated the turbulent transport coefficients $\alpha$ and $\eta_\mathrm{t}$ based on closure approaches. We found  an $\alpha$ quenching behaviour comparable to \citet{Brandenburg2008} and calculated the growth rates based on  $\alpha$ and $\eta_\mathrm{t}$. 
They appear to be in a
similar approximate range, but there is a number of differences when compared to the direct measurements. 
The difference between closure and direct estimates are a possible result of the fact that the former are very rough. To improve and check the goodness of the estimate, a more refined method such as the test-field method \citep{Schrinetal07} would be required, which is not currently supported by \textit{Astaroth}. In addition, in further studies the number of points on the $\Rem$ axis should be increased. 

As we are also used this study as a way for exploring the scientific potential of the \textit{Astaroth} API, some remarks should be made on the practical aspects of computation. The work presented here has benefited significantly from the performance enhancement provided by \textit{Astaroth}. It was feasible to perform the dynamo simulations and related tests with two computing nodes, with four Tesla P100 devices per node. This made our simulation very affordable within the limits of the ASIAA high-performance computing cluster. More discussion of GPU performance can be found in the Appendix \ref{sec:performance}. 
Additionally, it should be noted that an efficient GPU code can produce substantial amounts of data. 
Therefore, benefits of the performance will come in contact with the limitation of the data processing tools, that might not be as efficient and/or optimized as the GPU code. 
Astrophysicists rely on data analysis libraries, and we should note that for the maximal benefit of GPU performance, also connected data processing tools should be improved in efficiency.

For future enhancements to this work, there are several possibilities. First, more points in the $\Rem$-space could be covered. To better understand the scenario illustrated by our hypothesis depicted in Figure \ref{fig:LSDvsSSD}.
As the computation is efficient, this approach is basically limited by the available data storage. Second, a test-field method could be implemented to estimate the turbulent transfer coefficients  $\alpha$ and $\eta_\mathrm{t}$ in a more precise manner. 
The third possibility is to increase resolution and therefore $\Rem$ with the multi-node MPI implementation of \textit{Astaroth} becoming available.

\section{Conclusions}\label{sec:conclusion}

In this paper, we extended the \textit{Astaroth} library to work on multiple GPUs and applied it to study the turbulent dynamo problem.
Our implementation scaled from one to four GPUs with at least $90\%$ efficiency and exhibited a speedup of $35$  in single-node performance on four V100-SXM2-32GB GPUs compared with \textit{Pencil Code} runs on two 20-core Intel Xeon Gold 6230 Cascade Lake CPUs. Because of the limitations of CUDA peer-to-peer memory transfers, our implementation was restricted to a single node. Our results demonstrate that one-dimensional decomposition is sufficient to hide communication latencies within a node when carrying out computation and communication in parallel on current hardware. However, we expect that the use of MPI and multidimensional decomposition schemes are required for witnessing further scaling. 

We simulated helical and non-helical MHD turbulence with homogeneous random forcing, and by modifying  diffusivity and viscosity 
within the resolution limits
to investigate the dependence of the dynamo growth
on the magnetic Reynolds number, while keeping the magnetic Prandtl number fixed to unity.
We were able to extend the $\Rem$ range to
somewhat larger values than in some of the older studies, but most importantly, produce a large set of simulations to determine the dependence more accurately than before.
We estimated growth rates from the simulations and saw that with helical turbulence an LSD would 
grow
at any $\Rem\gtrsim 1$,  
while SSD would appear only beyond a critical value  $\Rem \sim 25$. SSD growth rates 
followed a clear logarithmic $\Rem$ dependence.
Earlier studies have either not been able to determine a clear dependency due to the small amount of data points, or reported consistency with a $\sqrt{\Rem}$ dependency.
In helical simulations with $\Rem >$ 25, both dynamo instabilities are evidently present simultaneously.
To further inspect and separate the signatures of the SSD and LSD,
we determined magnetic powerspectra.
 LSD spectra displayed growth at the largest scales while SSD ones showed a Kazantsev profile 
at low to intermediate wavenumbers, and peaked at scales smaller than the forcing scale.
Spectra in the regime where both dynamo instabilities act together retains characteristic of the LSD at  low wavenumbers, and those of the SSD at high wavenumbers,
although there is always a peak at the forcing scale.
In such circumstances, the powerspectra display SSD features during the exponential growth stage.
We computed probability density functions of the magnetic field, which showed
exponential shapes in the case of SSD, and 
a Gaussian distribution deformed by a top-hat profile from the mean (Beltrami) field for LSD. 
They also showed evidence for the coexistence of SSD
with LSD towards high $\Rem$.
To explain the behaviour of the growth rate in helically driven simulations we presented a hypothesis that the growth of LSD is inhibited by a budding SSD around medium $\Rem$.
We analyzed LSD
using closure estimates of turbulent transport coefficients. These  estimated 
growth rates agreed at low and medium $\Rem$ with the direct measurements, 
but diverged 
at high $\Rem$. The closure--estimated $\alpha$ displayed 
clear signs of quenching at high $\Rem$. 

\acknowledgments

\textit{Astaroth} is open source and available under GPL 3 license at \url{https://bitbucket.org/jpekkila/astaroth/}.

This work utilized tools developed by the CHARMS group and high-performance computing resources and cluster in ASIAA. This research has made use of {SAO/NASA} Astrophysics Data System. Additional compute resources for this work were provided by CSC -- IT Center for Science.

Authors thank Dr. Chun-Fan Liu 
for an useful insight.

J.P., M.J.K., and M.R.\ acknowledge the support of the Academy of Finland
ReSoLVE Centre of Excellence (grant number 307411).
This project has received funding from the European Research Council (ERC)
under the European Union's Horizon 2020 research and innovation
programme (Project UniSDyn, grant agreement n:o 818665).
M.V., H.S., and R.K acknowledge funding support for Theory within ASIAA from Academia Sinica.
H.S. acknowledges grant support from Ministry of Science and Technology (MoST)
in Taiwan through 105-2119-M-001-044-MY3, and 108-2112-M-001-009-. 

\appendix 

\section{Performance and scaling}\label{sec:performance}

We ran the benchmarks on a compute cluster consisting of a total of 80
SuperServer 1029GQ-TVRT nodes. Each node consisted of two Intel Xeon Gold 6230
Cascade Lake $20$-core processors running at $2.1$ GHz and four Tesla
V100-SXM2-32GB GV100GL (rev a1) GPUs running at $1.53$ GHz. Each GV100GL was
connected to the other three GPUs within a node via NVLink 2.0. Nodes were
connected in a fat tree network via dual-rail Mellanox ConnectX-6 InfiniBand
HDR100 MT28908 adapters, stated to provide an aggregate bandwidth of $23$ GiB/s
per adapter~\citep{mellanox-whitepaper}. Error-correcting codes (ECC) were
enabled on both the CPU and GPU memory systems. It should be noted that the
performance benchmarks were run on a different computing cluster with newer
hardware than what was available for the physics simulations.
\par
The library used for the tests is available at~\citep{Astaroth2020}, commit
\verb!e5dc5ca!. The code was compiled with GCC $8.3.0$ and CUDA toolkit
$10.1.168$. MPI implementation was provided by Mellanox HPC-X software toolkit
$2.5.0$ compiled with CUDA support. 
The single-GPU performance of Astaroth was analyzed in detail in previous
work~\citep{Pekkila2019}, where the performance was shown to be bound by cache
bandwidth. Execution speed was roughly six times lower than the idealized
theoretical upper limit.
\par
The grid consisted of a total of $512^{3}$ points in the computational domain
in the strong scaling tests, shown in Figure~\ref{fig:strong-scaling}. In weak
scaling tests, shown in Figure~\ref{fig:weak-scaling}, we used $N_x = N_y =
512$ and elongated the computational domain along the $z$-axis depending on the
number of devices $p$ to $N_z = 512p$. The tests were run on one to four Tesla
V100 GPUs on a single node using double precision. In our test cases, all
fields  were initialized to random
values within range $[0, 1]$ and the simulation was advanced using a constant
time step $\delta t = 1.19209\cdot10^{-7}$. We benchmarked the code by measuring
the running time of $100$ integration steps after $10$ warm-up steps and
reported the integration step time at the $90$th percentile. Forcing and upwinding
were disabled in the benchmarks as they have negligible effects on computing
performance and do not affect communication.
\begin{figure}
    \begin{minipage}{0.49\textwidth}
        \centering
        \begin{tikzpicture}
        \pgfplotsset{
            width = 0.80\textwidth
        }
          \begin{axis}[
            xlabel = Number of GPUs,
            ylabel = Time per step (ms),
            xmode=log,
            log basis x = 2,
            ymode=log,
            scale only axis,
            legend cell align = left,
            xticklabel=\pgfmathparse{2^\tick}\pgfmathprintnumber{\pgfmathresult},
            log ticks with fixed point,
            ytick = data,
        ]
            \addplot table[x index=0, y index=2,col sep=comma] {strong-scaling.csv};
            \addlegendentry{Ideal}
            \addplot table[x index=0, y index=1,col sep=comma] {strong-scaling.csv};
            \addlegendentry{Measured}
          \end{axis}
        \end{tikzpicture}
        \caption{Strong scaling.}
        \label{fig:strong-scaling}
    \end{minipage}
    \begin{minipage}{0.49\textwidth} 
        \centering
        \begin{tikzpicture}
        \pgfplotsset{
            width = 0.80\textwidth
        }
          \begin{axis}[
            xlabel = Number of GPUs,
            ylabel = Time per step (ms),
            xmode=log,
            log basis x = 2,
            scale only axis,
            legend cell align = left,
            legend pos = north west,
            xticklabel=\pgfmathparse{2^\tick}\pgfmathprintnumber{\pgfmathresult},
            log ticks with fixed point,
        ]
            \addplot table[x index=0, y index=2,col sep=comma] {weak-scaling.csv};
            \addlegendentry{Ideal}
            \addplot table[x index=0, y index=1,col sep=comma] {weak-scaling.csv};
            \addlegendentry{Measured}
          \end{axis}
        \end{tikzpicture}
        \caption{Weak scaling.}
        \label{fig:weak-scaling}
    \end{minipage}
\end{figure}

We compared single-node performance of \textit{Astaroth} to the
\textit{Pencil Code}~\citep{Pencil2020}. The \textit{Pencil Code} was built using the Intel Fortran Compiler version $19.0.4$ and
HPC-X MPI version $2.5.0$, using optimization level \verb!O3!. We benchmarked the \textit{Pencil Code} on the same compute node
as \textit{Astaroth}, using all the available $40$ CPU cores. We measured the performance of $23.17$ ns per integration step 
per grid point in a test case equivalent to the one used in the GPU benchmarks. The exact test case is available
at a dedicated code repository\footnote{\url{bitbucket.org/jpekkila/vaisala-pekkila-2020-artifacts}}. In contrast, the integration time per grid 
point with \textit{Astaroth} was $0.65$ ns on $4\times$ Tesla V100 GPUs. This gives us 
$35\times$ speedup with \textit{Astaroth} over the \textit{Pencil Code} in single-node performance.

The arithmetic performance and memory bandwidth of CPUs and GPUs can be used to
calculate a rough estimate for the speedup that can be gained by computing
data-parallel tasks on GPUs. In the ideal case, where both implementations
fully utilize the hardware resources, the GPU implementation can be expected to
exhibit a speedup of $12$--$13$  on the compute nodes used in this
work~\footnote{The aggregate performance on $4\times$ V100 GPUs is $31.32$
TFLOPS (arithmetic) and $3~452$ GiB/s (memory), whereas the respective numbers
for two Intel Xeon Gold 6230 CPUs are $2.5$ TFLOPS and $262$ GiB/s.}. We
measured the effective aggregate bandwidth to be $2~808$ GiB/s and $173$ GiB/s
on GPUs and CPUs, respectively. For bandwidth-bound kernels, this would give an
idealized speedup of $16$.

As the speedup with Astaroth was higher than expected, we suspect that the
\textit{Pencil Code} does not fully utilize the hardware, even though we strove
to employ the optimal build and runtime parameters for the benchmarks.
Optimizing the internal implementation of the \textit{Pencil Code} was out of
scope of this work.

\bibliographystyle{aasjournal}
\bibliography{references}

\end{document}